\pdfoutput=1
\documentclass[journal]{IEEEtran}
\usepackage{cite}

\usepackage{amsmath}

\usepackage{algpseudocode}

\ifCLASSINFOpdf
  \usepackage[pdftex]{graphicx}
\else
   \usepackage[dvips]{graphicx}
\fi
\ifCLASSOPTIONcompsoc
 \usepackage[caption=false,font=normalsize,labelfont=sf,textfont=sf]{subfig}
\else
\usepackage[caption=false,font=footnotesize]{subfig}
\begin{document}
%
\title{Convergence Analysis for Regular Wireless Consensus Networks}
%
%
%

\author{Sateeshkrishna Dhuli, Kumar Gaurav, and
        {Y.N.Singh,~\IEEEmembership{Senior Member,~IEEE}}
}

%
%

\markboth{}%
{Shell \MakeLowercase{\textit{et al.}}: Convergence Analysis for Regular Wireless Consensus Networks}
%



\maketitle

\begin{abstract}
Average consensus algorithms can be implemented over wireless sensor networks (WSN), where global statistics can be computed using communications among sensor nodes locally. Simple execution, robustness to global topology changes due to frequent node failures and underlying distributed philosophy have made consensus algorithms more suitable to WSNs. Since these algorithms are iterative in nature, it is very difficult to predict the convergence time of the average consensus algorithm on WSNs. We study the convergence of the average consensus algorithms for WSNs using distance regular graphs. We have obtained the analytical expressions for optimal consensus parameter and optimal convergence parameter which estimates the convergence time for $r$-nearest neighbor cycle and torus networks. We have also derived the generalized expression for optimal consensus parameter and optimal convergence parameter for $m$-dimensional $r$-nearest neighbor torus networks. The obtained analytical results agree with the simulation results and shown the effect of network dimension, number of nodes and nearest neighbors on convergence time. This work provides the basic analytical tools for managing and controlling the performance of average consensus algorithms over finite sized practical WSNs.
\end{abstract}


%
\IEEEpeerreviewmaketitle

\section{Introduction}

\IEEEPARstart{C}{onsensus} algorithms have received a lot of attention due to their ability to compute the desired global statistics by exchanging information only with direct neighbors. Average consensus algorithms have been extensively studied in distributed agreement and synchronization problems in the multi agent systems and load balancing in parallel computers ( \cite{algorithm}, \cite{loadbalance}, \cite{cooperation}). In contrast to centralized algorithms, the underlying distributed and decentralized philosophy avoids the need of any central fusion for collecting the information. This approach is particularly suitable in the following situations: 1) global network topology information is not known; 2) dynamic topology changes because of frequent node failures; 3) the nodes are computationally constrained and incapable to support the sophisticated routing techniques.

Distributed average consensus algorithms can be applied to WSNs for data fusion \cite{Robust}, \cite{fusion}, \cite{senran}. As the consensus algorithms are iterative in nature, convergence rate of the algorithms greatly influences the performance of the WSNs and it is lower bounded by the second smallest eigenvalue of the graph Laplacian \cite{fiedler}. To make this algorithm useful in a sensor network context, it is necessary to maximize the convergence rate to reach the consensus as soon as possible. Thus, extensive studies to increase the convergence rate have been done in the literature. In \cite{Distri}, distributed average consensus has been considered when the topology is random and the communication in the channels is corrupted by additive noise. It was proved that running the consensus for long time reduces the bias of the final average estimate but increases its variance. A closed form expression for the mean square error of the state and the optimum choice of parameters have been derived in \cite{Fastmean} to guarantee the fastest convergence. Consensus on small world and ramanujan networks has been studied in \cite{small}, \cite{ram} and it has been proved that convergence rate is maximized for these topologies. Optimal topology framework which increases the convergence rate and  minimizes the energy consumption has been studied in \cite{opttop}. In our work, we study the convergence of the consensus algorithm for finite distance-regular networks with varying number of nearest neighbors. These finite sized networks represent the notion of geographical proximity in the  practical WSNs. The main motivation for using the regular graph model is that most of the practical WSNs are finite sized which cannot be studied by asymptotic results existing in the literature. Random geometric graphs that model WSNs have similar asymptotic behavior as regular graph \cite{gossip}, which are convenient to analyze the wireless networks \cite{analysis}, \cite{stability}, \cite{reg}. Our results are more precise and can be applied to most of the practical WSNs. In $r$-nearest neighbor cycle and torus, an edge exists between every pair of neighbors that are $r$ hops away. If a node's transmission radius is increased, it will able to communicate with more number of nodes. Similarly, overhead increases with the number of nearest neighbors. So, we can consider the variable $r$ as both transmission radius and node overhead in WSNs. Effect of communication parameters on consensus algorithm's convergence rate has been studied analytically in \cite{vanka}. But it does not provide the exact formulation of the convergence time for average consensus algorithms. 
Despite the fact that distributed average consensus is simple to implement, it is generally difficult to predict its convergence time. Although there have been several studies of wireless consensus networks, analytical tools to control the network performance for WSNs are still inadequate. In this paper, we derive the generalized expressions to efficiently and exactly compute the optimal convergence, consensus parameter to estimate the convergence time for finite WSNs. This kind of analysis helps in estimating the convergence time efficiently as it avoids usage of computationally expensive algorithms which depends on thousands of simulation trails.
  
In summary, the paper is organized as follows. In Section II, we have given brief review of the consensus algorithms. In Section III, we have discussed the $r$-nearest neighbor networks and derived the  generalized eigenvalue expressions of weight matrix for the  $m$-dimensional WSNs. Analytical expressions for optimal consensus and optimal convergence parameters have been derived in Section IV. Finally, in Section V, simulation results have been presented and compared with the obtained analytical results. \\
%
\section{Review of consensus algorithms}
Let $G= (V,E)$, be an undirected graph with node set $V = \left\{ {1,2,......n} \right\}$ and an edge set $E \subseteq V \times V$. And, let $A$ be the $n\times n$ symmetric adjacency matrix of the graph $G$, where each entry of adjacency matrix is represented by $a_{ij}$, which is $1$ if node $i$ is connected to node $j$, else it is $0$. The degree matrix $D$ is defined as the diagonal matrix whose entry is $d_{ii}  = \deg (v_i )$, where $deg(\nu_{i})=\sum_{j=1}^{n}a_{ij}=\sum_{j=1}^{n}a_{ji}$. The Laplacian matrix of the graph $G$ is $n \times n$ symmetric matrix $L=D-A$, whose entries are

\begin{equation}
l_{ij}  = l_{ji}  = \left\{ \begin{array}{l}
 \deg (v_i )\,\,\,if\,\,j = i ,\\ 
  - a_{ij} \,\,\,\,\,\,\,\,\,if\,\,j \ne \,i .\\ 
 \end{array} \right.
 \label{1}
\end{equation}

The Laplacian is a positive definite matrix, hence the eigenvalues are given by
\begin{equation}
0=\lambda _{1}\left ( L \right )\leq \lambda _{2}\left ( L \right )\leq \lambda _{3}\left ( L \right )..........\leq \lambda _{N}\left ( L \right ).
\label{2}
\end{equation}

The graph topology is connected only if its zero eigenvalue has multiplicity one. The second smallest eigenvalue $\lambda _{2}\left ( L \right )> 0$ is the algebraic connectivity or the fiedler value \cite{fiedler} of the network.

Let $x_i (0)$ be the real scalar assigned to the node $i$ at $t=0$. Average consensus algorithm computes the average $x_{avg}=\frac{{\sum\nolimits_{i = 1}^n {x_i (0)} }}{n}$ at every node through a decentralized approach which does not require the sink node/base station. The minimization of disagreement between the $x_i (0)$ of the interacting nodes is expressed as a quadratic form \cite{god} on Laplacian.

\begin{equation}
\begin{array}{l}
 J(x):=\frac{1}{4}\sum_{i=1}^{n} \sum_{j\epsilon N_{i}}^{n}(x_{i}-x_{j})^{2}, \\ 
 \,\,\,\,\,\,\,\,\,\,\,\,\,\,\,\, =\frac{1}{4}\sum_{i=1}^{n} \sum_{j\epsilon N_{i}}^{n}a_{ij}(x_{i}-x_{j})^{2}, \\ 
\,\,\,\,\,\,\,\, \,\,\,\,\,\,\,\, =\frac{1}{2}x^{T}Lx.
 \end{array}
 \label{3}
\end{equation}

The quadratic form in (\ref{3}) is achieved using a simple steepest descent technique, and the minimum of (\ref{3}) updated by the \cite{cooperation} following expression.

\begin{equation}
\dot{x}(t)=-Lx(t).
\label{4}
\end{equation}
where $\dot{x}(t)$ denotes the average of all the real scalar values. 

At each step, node $i$ carries out its update based on its local state and communication with it's direct neighbors.
\begin{equation}
x_i (t + 1) = x_i (t) + h\sum\limits_{j \in N_i } {(x_j (t) - x_i (t))} ,\,\,\,i = 1,...,n,
\label{5}
\end{equation}
where $h$ is a consensus parameter and $N_i$ denotes neighbor set of node $i$.
This iterative method is expressed as the simple linear iteration
\begin{equation}
x(t + 1) = Wx(t),\,\,\,\,t = 0,1,2...,
\label{6}
\end{equation}

where $W$ denotes weight matrix, and $W_{ij}$ is a weight associated with the edge $(i,j)$. If we assign equal weight $h$ to each link in the network, then from \cite{Fast}, optimal weight for a given topology is

\begin{equation}
W_{ij}=\left\{\begin{matrix}
h & if ,\,\,\,\,(i,j) \in E,\\
1-hdeg(\nu_{i})&if,\,\,\,\,i=j,\\
0 & otherwise.
\end{matrix}\right.
\label{7}
\end{equation}
where optimal link weight or consensus parameter \cite{Fast} is defined as
\begin{equation}
 h  = \frac{2}{{\lambda _2 (L) + \lambda _{n} (L)}}.
 \label{8}
 \end{equation}
and Weight matrix is given by
\begin{equation}
W = I - hL.
\label{9}
\end{equation}
where $I$ is a $n\times n$ identity matrix. Let $\lambda _n (W)$ be the $n^{th}$ eigenvalue of $W$, then $\lambda _n (W)=1-h\lambda _n (L)$ satisfies
 \begin{equation}
1=\lambda _{1}(W)> \lambda _{2}(W)> \lambda _{3}(W).........\lambda _{n}(W).
\label{10}
\end{equation}
From (\ref{6}) convergence rate is expressed as
\begin{equation}
\left \| x_{i}-x_{avg} \right \|\leq \left \| x(0)-x_{avg} \right \| \gamma ^{i},
\label{11}
\end{equation}
where convergence parameter ($\gamma$) is
\begin{equation}
\gamma=\frac{\lambda_{n}(L)-\lambda_{2}(L)}{\lambda_{n}(L)+\lambda_{2}(L)}.
\label{12}
\end{equation}
\section{$r$-nearest neighbor networks}
Multi-dimensional geometric wireless network topologies can be represented by $r$-nearest neighbor networks \cite{analysis}, where nodes in the distance $r$ gets connected. The variable $r$ models the node's transmission radius or node overhead. These networks are one particular class of distance regular networks, where $r$-nearest neighbor cycle and $r$-nearest neighbor torus represents one dimensional and two dimensional topologies respectively.
\subsection{$r$-nearest neighbor cycle}
The $r$-nearest cycle can be represented by a circulant matrix \cite{Circul}, and it is given by
 \begin{equation}
\left[ \begin{array}{l}
 a_1 \,\,a_2 \,\,........a_{n - 1} \,\,a_n  \\ 
 a_n \,\,a_1 \,\,....\,....a_{n - 2} \,a_{n - 1}  \\ 
 .\,\,\,\,\,\,\,.\,\,\,\,\,\,\,\,\,\,\,\,\,\,\,\,\,\,\,\,\,\,\,\,.\,\,\,\,\,\,\,\,\,. \\ 
 .\,\,\,\,\,\,\,.\,\,\,\,\,\,\,\,\,\,\,\,\,\,\,\,\,\,\,\,\,\,\,\,.\,\,\,\,\,\,\,\,\,. \\ 
 a_3 \,\,a_4 \,\,\,...........a_1 \,\,\,a_2  \\ 
 a_2 \,\,a_3 \,\,.............a_n \,\,a_1 \,\, \\ 
 \end{array} \right].
 \label{13}
\end{equation}
where $a_{i}$, $i=1$ to $n$ represents the topology coefficients.
The $k^{th}$ eigenvalue of a circulant matrix is defined as
\begin{equation}
\lambda _k  = a_1  + a_2 \omega ^{k}  + .............. + a_n \omega ^{(n - 1)k}. 
\label{14}
\end{equation}
where $\omega$ is the $n^{th}$ root of 1, given by
\begin{equation}
\omega  = \cos \left( {\frac{{2\pi }}{n}} \right) + z\sin \left( {\frac{{2\pi }}{n}} \right)= e^\frac{2\pi z}{n}.
\label{15}
\end{equation}
The 1-nearest cycle and 2-nearest cycle are shown in Fig. 1 and Fig. 2 respectively. Then adjacency matrix ($A$) of a 1-nearest cycle is
\begin{equation}
A=\left[ \begin{array}{l}
 0\,\,1\,\,0\,\,..............0\,\,1 \\ 
 1\,\,0\,\,1\,\,..............0\,0 \\ 
 .\,\,\,\,.\,\,\,\,.\,\,\,\,\,\,\,\,\,\,\,\,\,\,\,\,\,\,\,\,\,\,\,\,\,.\,\,\,. \\ 
 .\,\,\,\,.\,\,\,\,.\,\,\,\,\,\,\,\,\,\,\,\,\,\,\,\,\,\,\,\,\,\,\,\,\,.\,\,\,. \\ 
 0\,\,0\,\,0\,\,..............0\,\,1 \\ 
 1\,\,0\,\,0\,\,..............1\,0\, \\ 
 \end{array} \right],
 \label{16}
\end{equation}
 and degree matrix ($D$) of a 1-nearest cycle is expressed as
\begin{equation}
D=\left[ \begin{array}{l}
 2\,\,0\,0\,\,............0\,\,0 \\ 
 0\,\,2\,0\,\,....\,........0\,\,0 \\ 
 .\,\,\,\,.\,\,\,\,\,\,\,\,\,\,\,\,\,\,\,\,\,\,\,\,\,\,\,\,\,\,\,\,\,.\,\,\,. \\ 
 .\,\,\,\,.\,\,\,\,\,\,\,\,\,\,\,\,\,\,\,\,\,\,\,\,\,\,\,\,\,\,\,\,\,.\,\,\,. \\ 
 0\,\,0\,\,\,...............2\,0\,\,\, \\ 
 0\,\,0\,\,\,...............0\,\,2 \\ 
 \end{array} \right].
 \label{17}
\end{equation}
$Theorem$ $1$: The $k^{th}$ eigenvalue of a weight matrix ($W$) for $1$-nearest neighbor cycle \cite{loadbalance} is
\begin{equation}
\resizebox{.7 \hsize} {!} {$\lambda _k (W) = \left( {1 - 2h} \right) + 2h {\cos \left( {\frac{{2\pi k}}{n}} \right)}$},
\label{18}
\end{equation}
where $k=0,1,...(n-1)$.\\
$Proof$: Using (\ref{16}) and (\ref{17}), Laplacian matrix for $1$-nearest neighbor cycle is expressed as
\begin{equation}
L=\left[ \begin{array}{l}
 \,\,2\,\,\,\,-1\,\,\,\,\,\,\,\,\,\,\,0\,\,.............0\,\,-1 \\ 
 -1\,\,\,\,\,\,\,\,2\,\,\,\,-1\,\,.............0\,\,\,\,\,\,\,\,\,0 \\ 
 \,\,\,.\,\,\,\,\,\,\,\,\,\,\,\,\,.\,\,\,\,\,\,\,\,\,\,\,\,.\,\,\,\,\,\,\,\,\,\,\,\,\,\,\,\,\,\,\,\,\,\,\,\,\,.\,\,\,\,\,\,\,\,\,\,. \\ 
\,\,\, .\,\,\,\,\,\,\,\,\,\,\,\,\,.\,\,\,\,\,\,\,\,\,\,\,\,.\,\,\,\,\,\,\,\,\,\,\,\,\,\,\,\,\,\,\,\,\,\,\,\,\,.\,\,\,\,\,\,\,\,\,\,. \\ 
\,\,\, 0\,\,\,\,\,\,\,\,\,\,\,0\,\,\,\,\,\,\,\,\,\,\,0\,\,.............2\,-1 \\ 
 -1\,\,\,\,\,\,\,\,\,0\,\,\,\,\,\,\,\,\,\,\,0\,\,.........-1\,\,\,\,\,\,\,\,2\, \\ 
 \end{array} \right].
 \label{19}
\end{equation}
Using (\ref{9}) and (\ref{19}), we get
\begin{equation}
\resizebox{.9 \hsize} {!} {$W=(I-Lh)=\left[ \begin{array}{l}
 (1-2h)\,\,\,\,\,\,\,\,\,\,\,h\,\,\,\,\,\,\,\,\,\,\,\,\,\,\,\,\,\,\,\,\,\,\,0\,\,................0\,\,\,\,\,\,\,\,\,\,\,\,\,\,\,\,\,\,\,\,\,\,\,h \\ 
\,\,\,\,\,\,\,\, h\,\,\,\,\,\,\,\,\,\,\,\,\,\,\,(1-2h)\,\,\,\,\,\,\,\,\,\,\,h\,\,................0\,\,\,\,\,\,\,\,\,\,\,\,\,\,\,\,\,\,\,\,\,\,\,0 \\ 
 \,\,\,\,\,\,\,\,\,\,\,\,\,.\,\,\,\,\,\,\,\,\,\,\,\,\,\,\,\,\,\,\,\,\,\,.\,\,\,\,\,\,\,\,\,\,\,\,\,\,\,\,\,\,\,\,\,\,.\,\,\,\,\,\,\,\,\,\,\,\,\,\,\,\,\,\,\,\,\,\,\,\,\,\,\,\,\,.\,\,\,\,\,\,\,\,\,\,\,\,\,\,\,\,\,\,\,\,\,\,\,\,\,. \\ 
 \,\,\,\,\,\,\,\,\,\,\,\,\,.\,\,\,\,\,\,\,\,\,\,\,\,\,\,\,\,\,\,\,\,\,\,.\,\,\,\,\,\,\,\,\,\,\,\,\,\,\,\,\,\,\,\,\,\,.\,\,\,\,\,\,\,\,\,\,\,\,\,\,\,\,\,\,\,\,\,\,\,\,\,\,\,\,\,.\,\,\,\,\,\,\,\,\,\,\,\,\,\,\,\,\,\,\,\,\,\,\,\,\,. \\ 
 \,\,\,\,\,\,\,\,\,\,\,0\,\,\,\,\,\,\,\,\,\,\,\,\,\,\,\,\,\,\,\,\,\,0\,\,\,\,\,\,\,\,\,\,\,\,\,\,\,\,\,\,\,\,\,0\,\,............(1-2h)\,\,\,\,\,\,\,\,\,\,h \\ 
 \,\,\,\,\,\,\,\,\,\,\,h\,\,\,\,\,\,\,\,\,\,\,\,\,\,\,\,\,\,\,\,\,\,0\,\,\,\,\,\,\,\,\,\,\,\,\,\,\,\,\,\,\,\,\,0\,\,...............h\,\,\,\,\,\,\,\,\,\,\,\,\,\,\,(1-2h)\, \\ 
 \end{array} \right]$}.
 \label{20}
\end{equation}
Finally, using (\ref{14}) and (\ref{20}), $k^{th}$ eigenvalue of $W$ is simplified as (\ref{18}).\\
$Theorem$ $2$: The $k^{th}$ eigenvalue of weight matrix ($W$) for $r$-nearest neighbor cycle is
\begin{equation}
\resizebox{.7 \hsize} {!} {$\lambda _k (W) = \left( {1 - 2rh} \right) + 2h\sum\limits_{j = 1}^r {\cos \left( {\frac{{2\pi kj}}{n}} \right)}$},
\label{21}
\end{equation}
where $k=0,1,...(n-1)$.\\
$Proof$: Using (\ref{14}), we observe that the first row is adequate to represent the eigenvalues of circulant matrix. 
The first row of Adjacency matrix (A) is
\begin{equation}
A_{1n}  = [0\,\underbrace {\,\,1\,\,\,1\,\,\,1\,\,}_{r\,times}\,.......0\,\,\,0.......\underbrace {1\,\,\,1\,\,\,1\,\,}_{r\,times}\,],
\label{22}
\end{equation} 
Similarly, first row of Degree matrix (D) is written as
\begin{equation}
D_{1n}  = \left[ {2r\,\,0\,\,0\,\,0\,.....0\,\,0\,\,0\,\,0} \right],
\label{23}
\end{equation}
So, using (\ref{22}) and (\ref{23}), first row of the Laplacian matrix (L) is expressed as
\begin{equation}
L_{1n}  = [2r\,\underbrace {\,\, - 1\,\,\, - 1\,\,\, - 1\,}_{r\,times}\,\,.......0\,\,\,0.......\underbrace { - 1\,\,\, - 1\,\,\, - 1\,\,}_{r\,times}\,],
\label{24}
\end{equation}
Using (\ref{24}) and (\ref{9}), first row of the $W$ for $r$-nearest neighbor cycle is expressed as
\begin{equation}
W_{1n}  = [(1 - 2rh)\,\,\,\underbrace {h\,\,\,h\,\,\,h}_{r\,times}\,\,\,.......0\,\,\,0.......\underbrace {\,\,\,h\,\,\,h\,\,\,h\,\,}_{r\,times}\,],
\label{25}
\end{equation}
Finally, using (\ref{14}) and (\ref{25}),  $k^{th}$ eigenvalue of $W$ for $r$-nearest neighbor cycle can be simplified as (\ref{21}).
\begin{figure}[!t]
\centering
\includegraphics[width=2in]{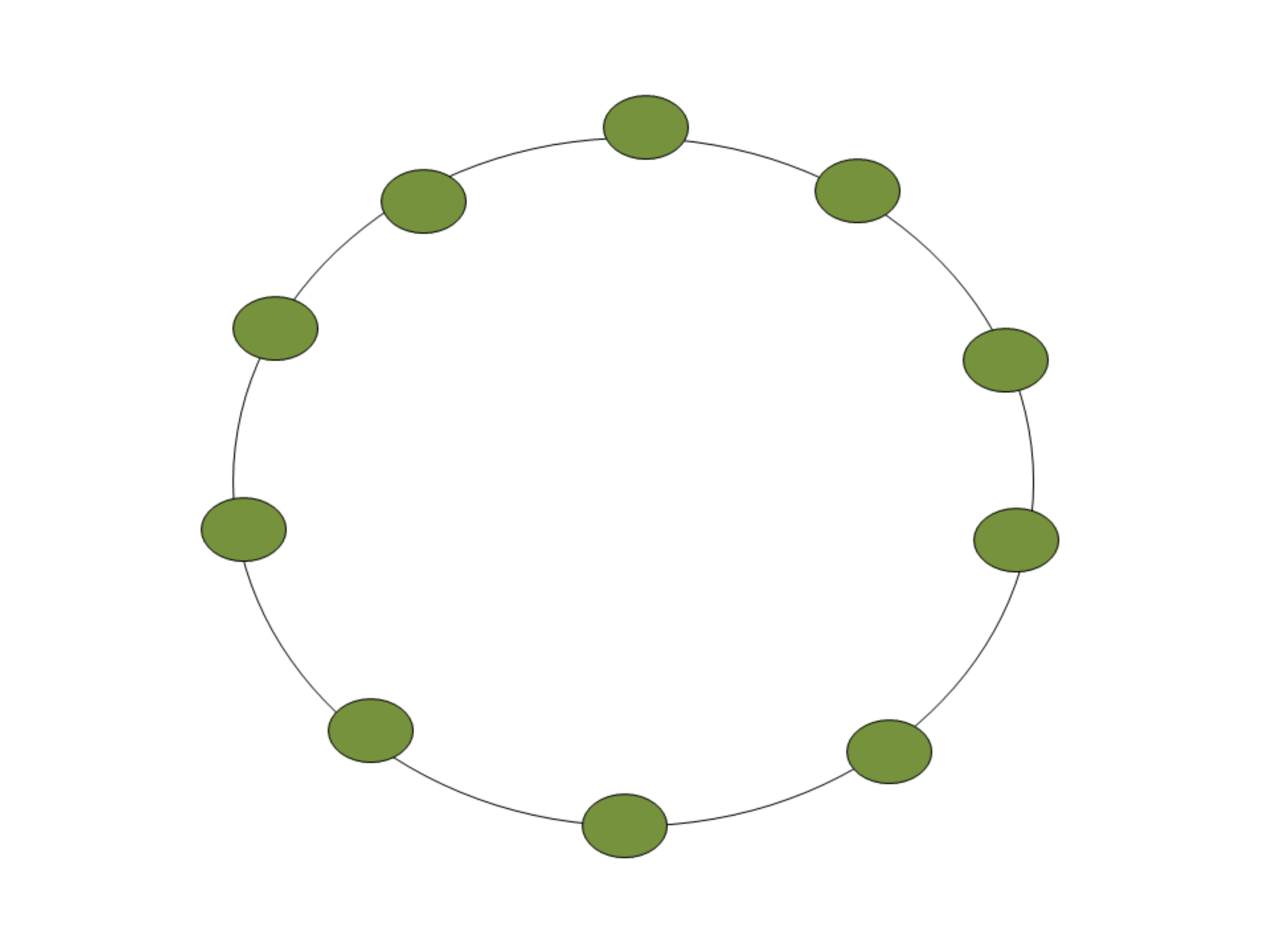}
\caption{1-nearest neighbor cycle.}
 \label{fig:1}
\end{figure}

\begin{figure}[!t]
\centering
\includegraphics[width=2in]{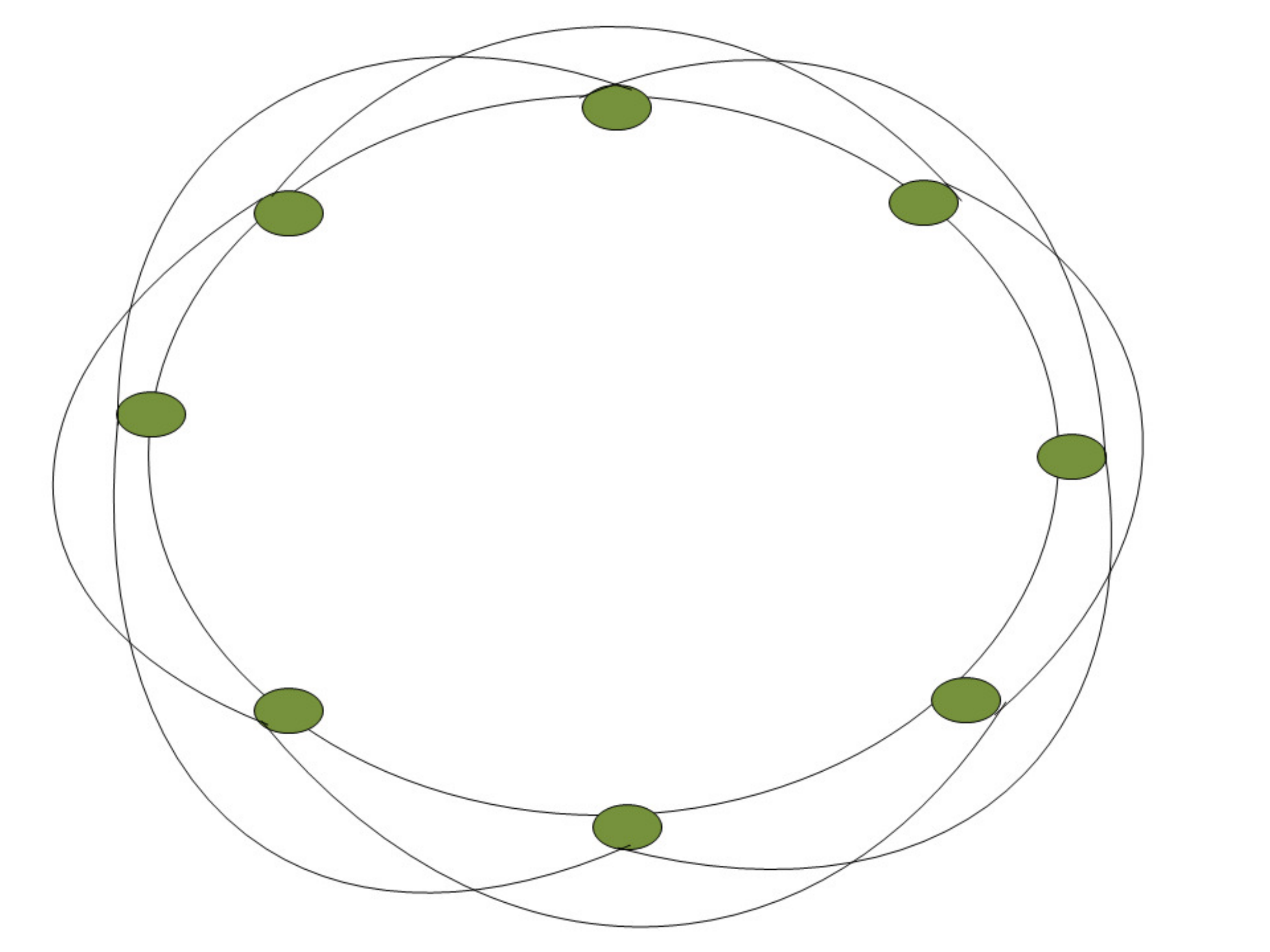}
\caption{2-nearest neighbor cycle.}
 \label{fig:1}
\end{figure}
\subsection{$r$-nearest neighbor torus}
A torus can be represented by $n \times n$ block circulant matrix $A$ as
\begin{equation}
A = \left[ \begin{array}{l}
 A_0 \,\,\,\,\,\,\,\,\,\,A_1 \,\,........A_{n_1 - 2} \,\,A_{n_1 -1}  \\ 
 A_{n_1 -1} \,\,A_0 \,\,....\,....A_{n_1 - 3} \,A_{n_1 - 2}  \\ 
 .\,\,\,\,\,\,\,\,\,\,\,\,\,\,\,\,\,\,\,\,\,.\,\,\,\,\,\,\,\,\,\,\,\,\,\,\,\,\,\,\,\,\,\,\,\,.\,\,\,\,\,\,\,\,\,\,\,\,\,\,\,. \\ 
 .\,\,\,\,\,\,\,\,\,\,\,\,\,\,\,\,\,\,\,\,\,.\,\,\,\,\,\,\,\,\,\,\,\,\,\,\,\,\,\,\,\,\,\,\,\,.\,\,\,\,\,\,\,\,\,\,\,\,\,\,\,. \\ 
 A_1 \,\,\,\,\,\,\,\,\,\,\,\,\,A_2 \,\,..........A_{n_1 -1} \,\,A_0 \,\, \\ 
 \end{array} \right],
 \label{26}
\end{equation}
Let the number of nodes $n=n_1^2$, then each block $A_i$, for $i=0,1...(n_1-1)$ represents $n_1 \times n_1$ circulant matrices.\\
Let $W_{k_1 ,k_2 }$ is a weight matrix for $k_1 \times k_2$ torus \cite{loadbalance}, then it is given by
\begin{equation}
\resizebox{.9 \hsize} {!} {$W_{k_1 ,k_2 }  = \left[ \begin{array}{l}
 (W_{k_1 }  - 2rhI_{k_1 }) \,\,\,\,\,hI_{k_1 } \,........\,........\,.......\,hI_{k1}  \\ 
 \,\,\,\,\,\,\,\,\,\,{hk1} \,\,\,\,\,\,\,\,\,\,(W_{k_1 }  - 2rhI_{k_1 }) \,\,hI_{k_1 } \,........\,.......\, \\ 
  \\ 
  \\ 
 \,........\,.......\,........\,.......\,\,hI_{k1} \,\,\,(W_{k_1 }  - 2rhI_{k_1 })  \\ 
 \end{array} \right]_{k_1  \times k_2 }$}.
 \label{27}
\end{equation}
where $W_{k_1}$ is the weight matrix of $r$-nearest neighbor cycle consists of $k_1$ number of nodes and $I_{k_1}$ is the identity matrix of order $k_1 \times k_1$.\\
$Theorem$ $3$: The eigenvalue $\lambda _{j_1 ,j_2 }$ of $W_{k_1 ,k_2 }$ of $1$-nearest neighbor torus \cite{loadbalance} is
\begin{equation}
1 - 4h + 2h\cos \left( {\frac{{2\pi j_1 }}{{k_1 }}} \right) + 2h\cos \left( {\frac{{2\pi j_2 }}{{k_2 }}} \right),
\label{28}
\end{equation}
where $j_1  = 0,1,2,...(k_1  - 1)$, $j_2  = 0,1,2,...(k_2  - 1)$.\\
$Proof$ :  
Using (\ref{14}) and (\ref{27}), the eigenvalue expression for $W_{k_1 ,k_2 }$ is
\begin{align}
\lambda _{j_1 ,j_2 } (W_{k_1 ,k_2 }) = \lambda \left( {W_{k_1 } } \right) - 2h + 2h\cos \left( {\frac{{2\pi j_2 }}{{k_2 }}} \right).\label{29}
\end{align}
Finally, substitution of (\ref{18}) in (\ref{29}) results in (\ref{28}).\\
$Theorem$ $4$: The eigenvalue $\lambda _{j_1 ,j_2 }$ of $W_{k_1 ,k_2 }$ for $r$-nearest neighbor torus is
\begin{equation}
\resizebox{.9 \hsize} {!} {$\lambda _{j_1 ,j_2 } (W_{k_1 ,k_2 })= (1 - 4rh) + 2h\sum\limits_{i = 1}^r {\cos \left( {\frac{{2\pi j_1 i}}{{k_1 }}} \right)}  + 2h\sum\limits_{i = 1}^r {\cos \left( {\frac{{2\pi j_2 i}}{{k_2 }}} \right)}$}
\label{30},
\end{equation}
where $j_1  = 0,1,2,...(k_1  - 1), j_2  = 0,1,2,...(k_2  - 1)$.\\
$Proof$ : The first row of weight matrix for $r$-nearest neighbor torus is
\begin{equation}
W_{1n}  = \left[ {W_{k_1 }  - 2rhI_{k_1 } \,\,\,\,\underbrace {hI_{k_1 } \,hI_{k_1 } .............hI_{k_1 } \,\,hI_{k_1 } }_{2r\,\,times}\,\,} \right],
\label{31}
\end{equation}
Using (\ref{14}) and (\ref{31}), we obtain
\begin{equation}
\lambda \left( {W_{k_1 } } \right) - 2rh + 2rh\sum\limits_{i = 1}^r {\cos \left( {\frac{{2\pi ij_2 }}{{k_2 }}} \right)},
\label{32}
\end{equation}
Therefore, substitution of (\ref{21}) in (\ref{32}) results in (\ref{30}).\\
$Theorem$ $5$: The eigenvalues $\lambda _{j_1 ,j_2 .....j_m }$ of weight matrix $W_{k_1 ,k_2 ,....k_m }$ for $m$-dimensional $r$-nearest neighbor torus is
\begin{equation}
\lambda _{j_1 ,j_2 .....j_m } (W) = (1 - 2mrh) + 2h\sum\limits_{j = 1}^r {\sum\limits_{i = 1}^m {\cos \left( {\frac{{2\pi j_i}}{{k_i }}} \right)} },
\label{33}
\end{equation}
where  $j_i  = 0,1,2,...(k_i  - 1)$.\\
$Proof$: 
From $Theorem$ $4$, the eigenvalue expression for two dimensional $r$-nearest neighbor torus is
\begin{equation}
\resizebox{.9 \hsize} {!} {$\lambda _{j_1 ,j_2 } (W_{k_1 ,k_2 })= (1 - 4rh) + 2h\sum\limits_{i = 1}^r {\cos \left( {\frac{{2\pi j_1 i}}{{k_1 }}} \right)}  + 2h\sum\limits_{i = 1}^r {\cos \left( {\frac{{2\pi j_2 i}}{{k_2 }}} \right)}$},
\label{34}
\end{equation}
Similarly, eigenvalue expression for three dimensional $r$-nearest neighbor torus is expressed as
\begin{equation}
\resizebox{1 \hsize} {!} {$
\lambda _{j_1 ,j_2 , j_3} (W_{k_1 ,k_2, k_3})\left( {1 - 6rh} \right) + 2rh\sum\limits_{i = 1}^r {\cos \left( {\frac{{2\pi j_1 i}}{{k_1 }}} \right)}  + 2rh\sum\limits_{i = 1}^r {\cos \left( {\frac{{2\pi j_2 i}}{{k_2 }}} \right)}  + 2rh\sum\limits_{i = 1}^r {\cos \left( {\frac{{2\pi j_3 i}}{{k_3 }}} \right)} $},
\label{35}
\end{equation}
Hence, without loss of generality, using (\ref{34}) and (\ref{35}), the eigenvalue expression of $m$-dimensional $r$-nearest neighbor torus can be written as (\ref{33}).\\
A two dimensional nearest neighbor torus can be defined in various ways as shown in the Fig. 3 and Fig. 4. To get the nearest neighbors in two dimensions, $L^1$ norm, $L^2$ norm, and $L^\infty$ norm are used. For simple and convenient analysis \cite{analysis}, $L^1$ norm and $L^\infty$ norm are preferable, where  $L^1$ norm is used when there is an connection between nodes whose shortest path is within $r$ hops on the torus and $L^\infty$ norm is used when vertical and horizontal distance are both within $r$ hops on the torus.  
\begin{figure}[!t]
\centering
\includegraphics[width=2 in]{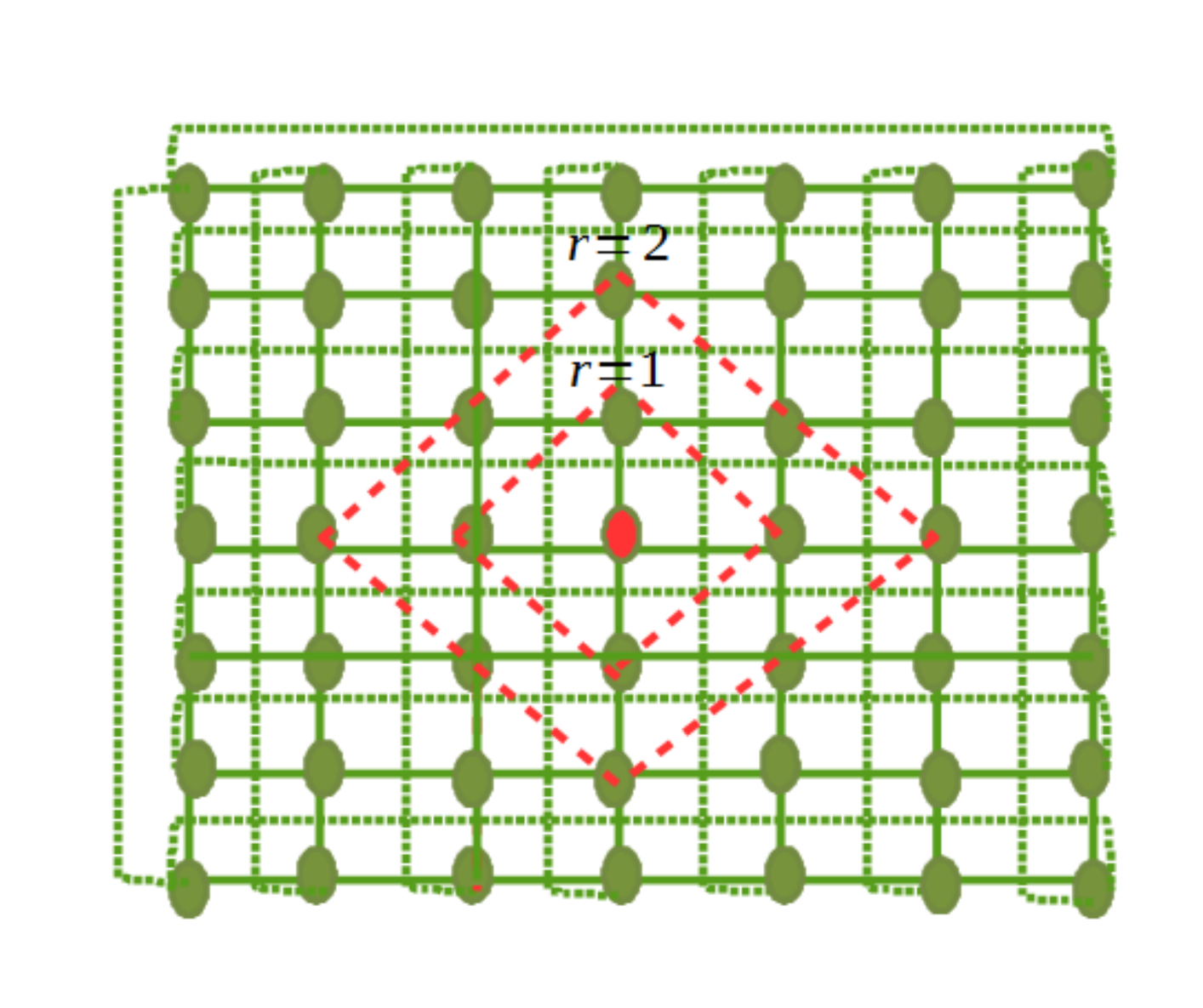}
\caption{$L^1$ Nearest neighbor torus.}
\label{fig:3}
\end{figure}
\begin{figure}[!t]
\centering
\includegraphics[width=2 in]{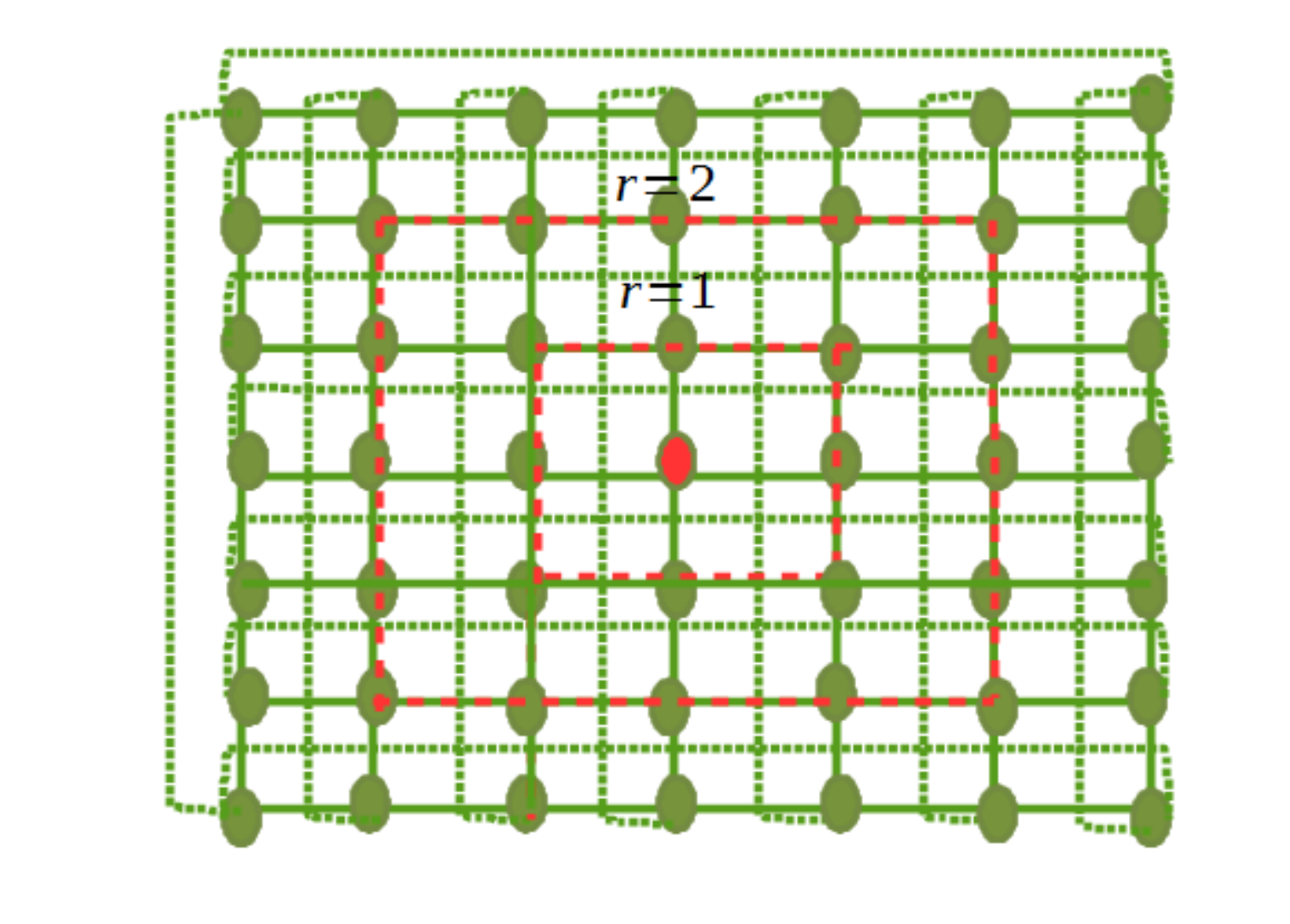}
\caption{$L^\infty$ Nearest neighbor torus.}
\label{fig:3}
\end{figure}
\section{Convergence analysis for $r$-nearest neighbor networks}
After calculating the eigenvalue expressions, the steps for calculating the optimal consensus parameter $h$, optimal convergence parameter $\gamma$, and convergence time $T$ are as follows:\\
(1) Observe the second largest and smallest eigenvalues of $W$.\\
(2) Calculate $h$ using $\lambda _2 (W) = -\lambda _n (W)$, here $\lambda _2 (W)$ is a second largest eigenvalue of $W$ and
 $\lambda _n (W)$ is a smallest eigenvalue of $W$.\\
(3) Determine $\gamma (W) = max\left\{ {\left| {1 - h\lambda _2 (L)} \right|,\left| {1 - h\lambda _n (L)} \right|} \right\}$.\\
(4) Finally, calculate the convergence time using $T = 1/\left( {\log (1/\gamma )} \right)$ \cite{step}.
 \subsection{$r$-nearest neighbor cycle}
$Theorem$ $6$: Given an $r$-nearest neighbor cycle $C_n^r $ and $n$ is even, the optimal consensus parameter $h$ is computed by
\begin{equation}
h = \frac{1}{{2r + 1 - \frac{1}{2}\left( {\frac{{\sin \left( {\frac{{2\pi \left( {r + 0.5} \right)}}{n}} \right)}}{{\sin \frac{\pi }{n}}} + \cos (\pi r)} \right)}}.
\label{36}
\end{equation}
$Proof$ : The proof is deferred to Appendix A.\\

$Theorem$ $7$: Given an $r$-nearest neighbor cycle $C_n^r $ and $n$ is odd, the optimal consensus parameter $h$ is computed by
\begin{equation}
\resizebox{.9 \hsize} {!} {$h = \frac{1}{{2r + 1 - \frac{1}{2}\left( {\frac{{\sin \left( {\frac{{(2r + 1)\pi }}{n}} \right)}}{{\sin \frac{\pi }{n}}} - \frac{{\cos \left( {\frac{{\pi (2r + 1)}}{{2n}}} \right)}}{{\cos \frac{\pi }{{2n}}}}} \right)}}$}.
\label{37}
\end{equation}
$Proof$ : The proof is deferred to Appendix A.\\

$Theorem$ $8$: Given an $r$-nearest neighbor cycle $C_n^r $ and $n$ is even, the optimal convergence parameter $\gamma$ is computed by
\begin{equation}
\gamma  = \frac{{\left( {\frac{{\sin \left( {\frac{{(2r + 1)\pi }}{n}} \right)}}{{\sin \frac{\pi }{n}}} - \cos (\pi r)} \right)}}{{4r + 2 - \left( {\frac{{\sin \left( {\frac{{(2r + 1)\pi }}{n}} \right)}}{{\sin \frac{\pi }{n}}} + \cos (\pi r)} \right)}}.
\label{38}
\end{equation}
$Proof$ : The proof is deferred to Appendix A.\\

$Theorem$ $9$: Given an $r$-nearest neighbor cycle $C_n^r $ and $n$ is odd, the optimal convergence parameter $\gamma$ is
\begin{equation}
\resizebox{.9 \hsize} {!} {$\gamma  = \frac{{\left( {\frac{{\sin \left( {\frac{{(2r + 1)\pi }}{n}} \right)}}{{\sin \frac{\pi }{n}}} + \frac{{\cos \left( {\frac{{\pi (2r + 1)}}{{2n}}} \right)}}{{\cos \frac{\pi }{{2n}}}}} \right)}}{{4r + 2 - \left( {\frac{{\sin \left( {\frac{{(2r + 1)\pi }}{n}} \right)}}{{\sin \frac{\pi }{n}}} - \frac{{\cos \left( {\frac{{\pi (2r + 1)}}{{2n}}} \right)}}{{\cos \frac{\pi }{{2n}}}}} \right)}}$}.
\label{39}
\end{equation}
$Proof$ : The proof is deferred to Appendix A.\\

\subsection{$r$-nearest neighbor torus}
$Theorem$ $10$: Given an $r$-nearest neighbor torus $T_{k_1 ,k_2 }^r $ and $k_1, k_2$ are even integers, the optimal consensus parameter is 
\begin{equation}
h = \frac{1}{{1.5 + 3r - \frac{1}{2}\left( {\frac{{\sin \left( {\frac{{(2r + 1)\pi }}{{k_2 }}} \right)}}{{\sin \frac{\pi }{{k_2 }}}} + 2\cos (\pi r)} \right)}}.
\label{40}
\end{equation}
$Proof$ : The proof is deferred to Appendix B.\\

$Theorem$ $11$: Given an $r$-nearest neighbor torus $T_{k_1 ,k_2 }^r $ and $k_1, k_2$ are odd integers, the optimal consensus parameter is expressed as
 \begin{equation}
\resizebox{.9 \hsize} {!} {$ h = \frac{1}{{1.5 + 3r- 0.5\left( {\frac{{\sin \left( {\frac{{(2r + 1)\pi }}{{k_2 }}} \right)}}{{\sin \left( {\frac{\pi }{{k_2 }}} \right)}} + \frac{{\sin \left( {\frac{{\pi (2r + 1)(k_1  - 1)}}{{2k_1 }}} \right)}}{{\sin \left( {\frac{{\pi (k_1  - 1)}}{{2k_1 }}} \right)}} + \frac{{\sin \left( {\frac{{\pi (2r + 1)(k_2  - 1)}}{{2k_2 }}} \right)}}{{\sin \left( {\frac{{\pi (k_2  - 1)}}{{2k_2 }}} \right)}}} \right)}}$}.
\label{41}
\end{equation}
$Proof$ : The proof is deferred to Appendix B.\\

$Theorem$ $12$: Given an $r$-nearest neighbor torus $T_{k_1 ,k_2 }^r $ and $k_1, k_2$ are even integers, the optimal convergence parameter is 
\begin{equation}
\gamma  = \frac{{r + 0.5 + 0.5\left( {\frac{{\sin \left( {\frac{{(2r + 1)\pi }}{{k_2 }}} \right)}}{{\sin \left( {\frac{\pi }{{k_2 }}} \right)}} - 2\cos \pi r} \right)}}{{(1.5 + 3r) - 0.5\left( {\frac{{\sin \left( {\frac{{(2r + 1)\pi }}{{k_2 }}} \right)}}{{\sin \left( {\frac{\pi }{{k_2 }}} \right)}} + 2\cos \pi r} \right)}}.
\label{42}
\end{equation}
$Proof$ : The proof is deferred to Appendix B.\\
$Theorem$ $13$: Given an $r$-nearest neighbor torus $T_{k_1 ,k_2 }^r $ and $k_1, k_2$ are odd integers, the optimal convergence parameter is 
 \begin{equation}
\resizebox{.9 \hsize} {!} {$\gamma  = \frac{{r + 0.5 + 0.5\left( {\frac{{\sin \left( {\frac{{\pi (2r + 1)(k_1  - 1)}}{{2k_1 }}} \right)}}{{\sin \left( {\frac{{\pi (k_1  - 1)}}{{2k_1 }}} \right)}} - \frac{{\sin \left( {\frac{{(2r + 1)\pi }}{{k_2 }}} \right)}}{{\sin \left( {\frac{\pi }{{k_2 }}} \right)}} + \frac{{\sin \left( {\frac{{\pi (2r + 1)(k_2  - 1)}}{{2k_2 }}} \right)}}{{\sin \left( {\frac{{\pi (k_2  - 1)}}{{2k_2 }}} \right)}}} \right)}}{{(1.5 + 3r) - 0.5\left( {\frac{{\sin \left( {\frac{{\pi (2r + 1)(k_1  - 1)}}{{2k_1 }}} \right)}}{{\sin \left( {\frac{{\pi (k_1  - 1)}}{{2k_1 }}} \right)}} + \frac{{\sin \left( {\frac{{(2r + 1)\pi }}{{k_2 }}} \right)}}{{\sin \left( {\frac{\pi }{{k_2 }}} \right)}} + \frac{{\sin \left( {\frac{{\pi (2r + 1)(k_2  - 1)}}{{2k_2 }}} \right)}}{{\sin \left( {\frac{{\pi (k_2  - 1)}}{{2k_2 }}} \right)}}} \right)}}$}.
\label{43}
\end{equation}
$Proof$ : The proof is deferred to Appendix B.\\
Most of the WSN applications, such as space monitoring, cave monitoring and studying underwater eco system operates in multiple dimensions. So without loss of generality, we have also derived the expressions for $h$ and $\gamma$ of $m$-dimensional torus networks. 
\subsection{$m$-dimensional torus}
$Theorem$ $14$: Given a $m$-dimensional $r$-nearest neighbor torus and $k_1, k_2, k_3...k_m$ are even integers, the optimal consensus parameter is computed by
\begin{equation}
h = \frac{1}{{(m + 1)(r + 0.5) - \frac{{0.5\sin \left( {\frac{{(2r + 1)\pi }}{{k_1 }}} \right)}}{{\sin \left( {\frac{\pi }{{k_1 }}} \right)}} - \frac{{m\cos \pi r}}{2}}}.
\label{44}
\end{equation}
$Proof$ : The proof is deferred to Appendix C.\\

$Theorem$ $15$: Given a $m$-dimensional $r$-nearest neighbor torus and $k_1, k_2, k_3...k_m$ are odd integers, the optimal consensus parameter is computed by
\begin{equation}
\resizebox{.9 \hsize} {!} {$h = \frac{1}{{(m + 1)(r + 0.5) - \frac{{0.5\sin \left( {\frac{{(2r + 1)\pi }}{{k_1 }}} \right)}}{{\sin \left( {\frac{\pi }{{k_1 }}} \right)}} - \sum\limits_{l = 1}^m {\frac{{0.5\sin \left( {\frac{{(2r + 1)\pi (k_l  - 1)}}{{2k_l }}} \right)}}{{\sin \left( {\frac{{\pi (k_l  - 1)}}{{2k_l }}} \right)}}} }}$}.
\label{45}
\end{equation}
$Proof$ : The proof is deferred to Appendix C.\\
$Theorem$ $16$: Given a $m$-dimensional $r$-nearest neighbor torus and $k_1, k_2, k_3...k_m$ are even integers, the optimal convergence parameter is computed by
\begin{equation}
\resizebox{.9 \hsize} {!} {$\gamma  = \frac{{(m - 1)(r + 0.5) + \frac{{0.5\sin \left( {\frac{{(2r + 1)\pi }}{{k_1 }}} \right)}}{{\sin \left( {\frac{\pi }{{k_1 }}} \right)}} - \frac{m}{2}\cos (\pi r)}}{{(m + 1)(r + 0.5) - \frac{{0.5\sin \left( {\frac{{(2r + 1)\pi }}{{k_1 }}} \right)}}{{\sin \left( {\frac{\pi }{{k_1 }}} \right)}} - \frac{m}{2}\cos (\pi r)}}$}.
\label{46}
\end{equation}
$Proof$ : The proof is deferred to Appendix C.\\
$Theorem  17$: Given a $m$-dimensional $r$-nearest neighbor torus and $k_1, k_2, k_3...k_m$ are odd integers, the optimal convergence parameter is computed by
 \begin{equation}
\resizebox{.9 \hsize} {!} {$\gamma  = \frac{{(m - 1)(r + 0.5) + \frac{1}{2}\left( {\frac{{\sin \frac{{(r + 0.5)2\pi }}{{k_1 }}}}{{\sin \frac{\pi }{{k_1 }}}} - \sum\limits_{l = 1}^m {\frac{{\sin \frac{{(r + 0.5)\pi (k_l  - 1)}}{{k_l }}}}{{\sin \frac{{\pi (k_l  - 1)}}{{2k_l }}}}} } \right)}}{{(m + 1)(r + 0.5) - \frac{1}{2}\left( {\frac{{\sin \frac{{(r + 0.5)2\pi }}{{k_1 }}}}{{\sin \frac{\pi }{{k_1 }}}} + \sum\limits_{l = 1}^m {\frac{{\sin \frac{{(r + 0.5)\pi (k_l  - 1)}}{{k_l }}}}{{\sin \frac{{\pi (k_l  - 1)}}{{2k_l }}}}} } \right)}}$}.
\label{47}
\end{equation}
$Proof$ : The proof is deferred to Appendix C.\\
\section{Numerical results and Discussion}
In this section, we present numerical results to examine the effect of $n$, $m$, and $r$ on the $h$, $\gamma$, and $T$. Plots of the $h$ versus $n$ for 1-nearest neighbor cycle are shown in Fig. 5. We have observed that $h$ increases with $n$ for small scale systems,and from $n=40$, optimal weight approaches approximately $0.5$. Plots of $\gamma$ versus $n$ and $T$ versus $n$ for $r=1$, are shown in Fig. 6 and Fig. 7 respectively. Convergence time $T$ values have been calculated using (\ref{38}) and (\ref{39}), for Fig. 7. As $n$ increases, $\gamma$ increases and approaches unity for large scale systems as shown in Fig. 6. From Fig. 7, we can observe that $T$ increases exponentially with $n$. 

Figs. 8, 9, and 10 show plots of the $h$, $\gamma$, and $T$ versus $r$ for $n=400$ respectively. From Fig. 8, We find that $h$ decreases exponentially with $r$ and $\gamma$ decreases linearly with $r$. From Fig. 10, we can see that $T$ decreases exponentially with $r$, since the links or edges between every node increases with $r$ directly influence the consensus process which in turn decreases the convergence time. Note that, from $r=10$, $T$ values are almost constant irrespective of increase in $r$ values. Figs. 11, 12, and 13 show plots of the $h$, $\gamma$, and $T$ versus $k_{1}$ and $k_{2}$ for respectively. We have used the (\ref{40}), (\ref{41}), (\ref{42}), and (\ref{43}) to compute the $h$, $\gamma$, and $T$ values for $r=1$. It is observed that optimal weight $h$ increases and approaches $0.25$ in two dimensional WSNs for $r=1$. As shown in Figs. 12 and 13, $\gamma$ increases with $k_{1}$ and $k_{2}$ and approaches unity and $T$ increases compared to Fig.7, because of the increase in total number of nodes in torus network.  

Figs. 14, 15, and 16 show plots of the $h$, $\gamma$, and $T$ versus $r$ respectively. For ease of analysis, we have assumed that  $k_1=1000$ and $k_2=1000$, and observe that $h$, $\gamma$, and $T$ decreases with $r$. Plots of $h$, $\gamma$, and $T$ versus $m$ with varying number of $r$ values are shown in Figs. 17, 18, and 19 respectively. In this case, we have assumed that $k_{1}$ = 16, $k_{2}$ = 18, $k_{3}$ = 20, $k_{4}$ = 22, $k_{5}$ = 24, and $k_{6}$ = 26. As shown in the Fig. 17, optimal weight $h$ decreases with $m$ for various $r$ values and a drastic decrease can be observed from $r=2$ to $r=3$ and $r=4$ to $r=5$. As shown in the Fig. 18, $\gamma$ values approaches different values which are less than unity depends on $m$ and $r$ values. Fig. 19 shows that convergence time $T$ increases linearly with $m$, but the increase in $r$ results in substantial decay of $T$ values. 
\subsection{Convergence time-overhead Optimization}
Using the expressions derived in $Theorem$ $8$, $Theorem$ $9$, $Theorem$ $12$, $Theorem$ $13$, $Theorem$ $16$, and $Theorem$ $17$, we can estimate the trade-off between convergence time and overhead. From the Figs. 10, 16, and 19, we can observe that nearest neighbors or node transmission radius is exponentially reducing the convergence time, which is a primary objective of consensus algorithm. But the node's power consumption \cite{vanka} is 
\begin{equation}
P=\left ( \frac{r}{\sqrt{n}} \right )^{\alpha}
\label{48}
\end{equation}

where $\alpha$ is a path-loss exponent. So it has to be noted that while applying consensus algorithms on WSNs, it is also necessary to take care about the node's power consumption since sensor nodes consist of limited power resources. We propose an optimization framework, to minimize the convergence time $T$ subject to total power consumption constraint and minimizing the power consumption subject to maximum convergence time. 

\begin{equation*}
\begin{aligned}
&{\text{minimize}}
& & T \\
& \text{subject to}
& & r\leq r_{max}, \; P\leq P_{max}
\end{aligned}
\label{49}
\end{equation*}

\begin{equation*}
\begin{aligned}
&{\text{minimize}}
& & P \\
& \text{subject to}
& & T\leq T_{max}, \; r\leq r_{max}
\end{aligned}
\label{50}
\end{equation*}
where $r_{max}$, $T_{max}$, and $P_{max}$ are certain threshold values defined based on WSN resource requirements. The solutions can be obtained by calculating the respective Lagrangian values.
\begin{figure}[!t]
\centering
\includegraphics[width=8 cm, height=4 cm]{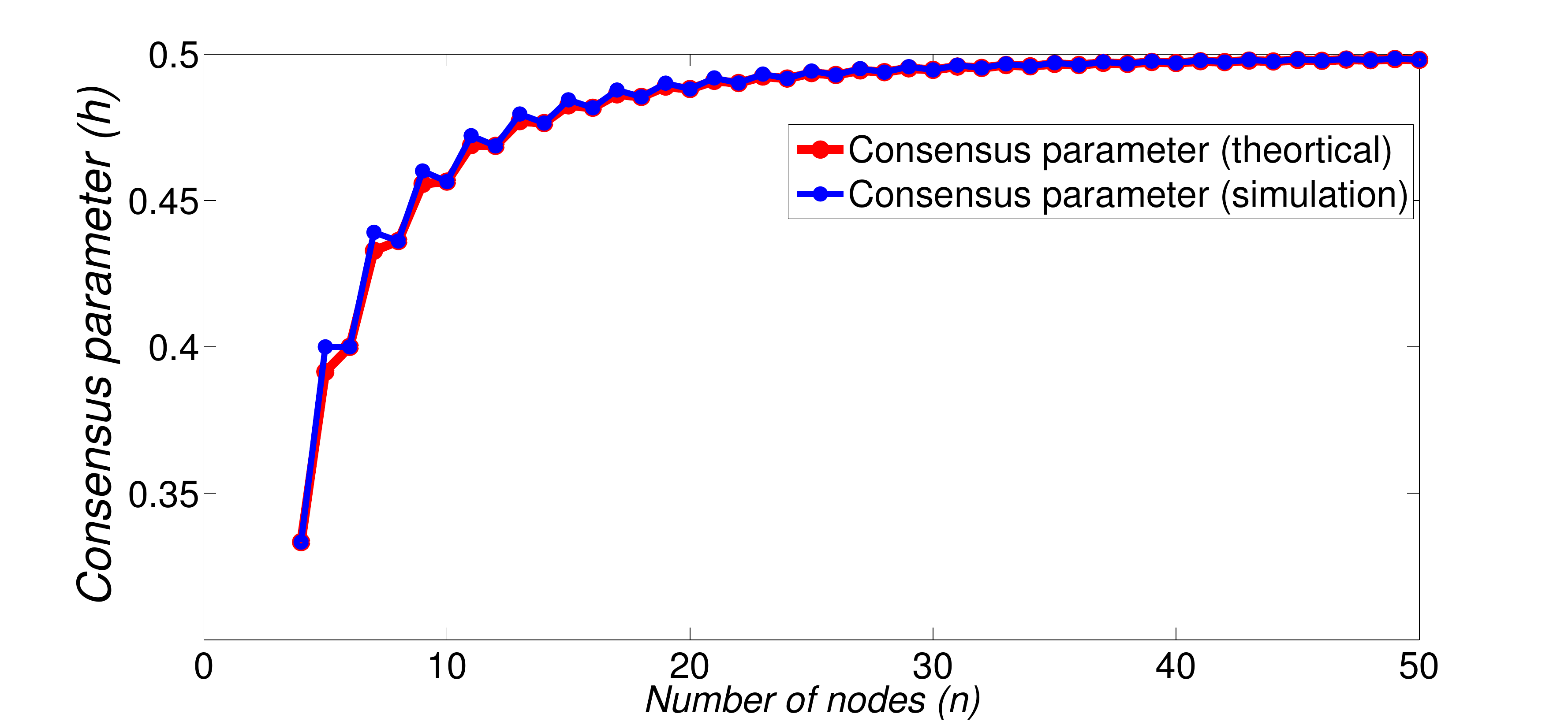}
\caption{Comparison of theoretical and simulation results for consensus parameter $h$ of $1$-nearest neighbor cycle.}
\label{fig:4}
\end{figure}
\begin{figure}[!t]
\centering
\includegraphics[width=8 cm, height=4 cm]{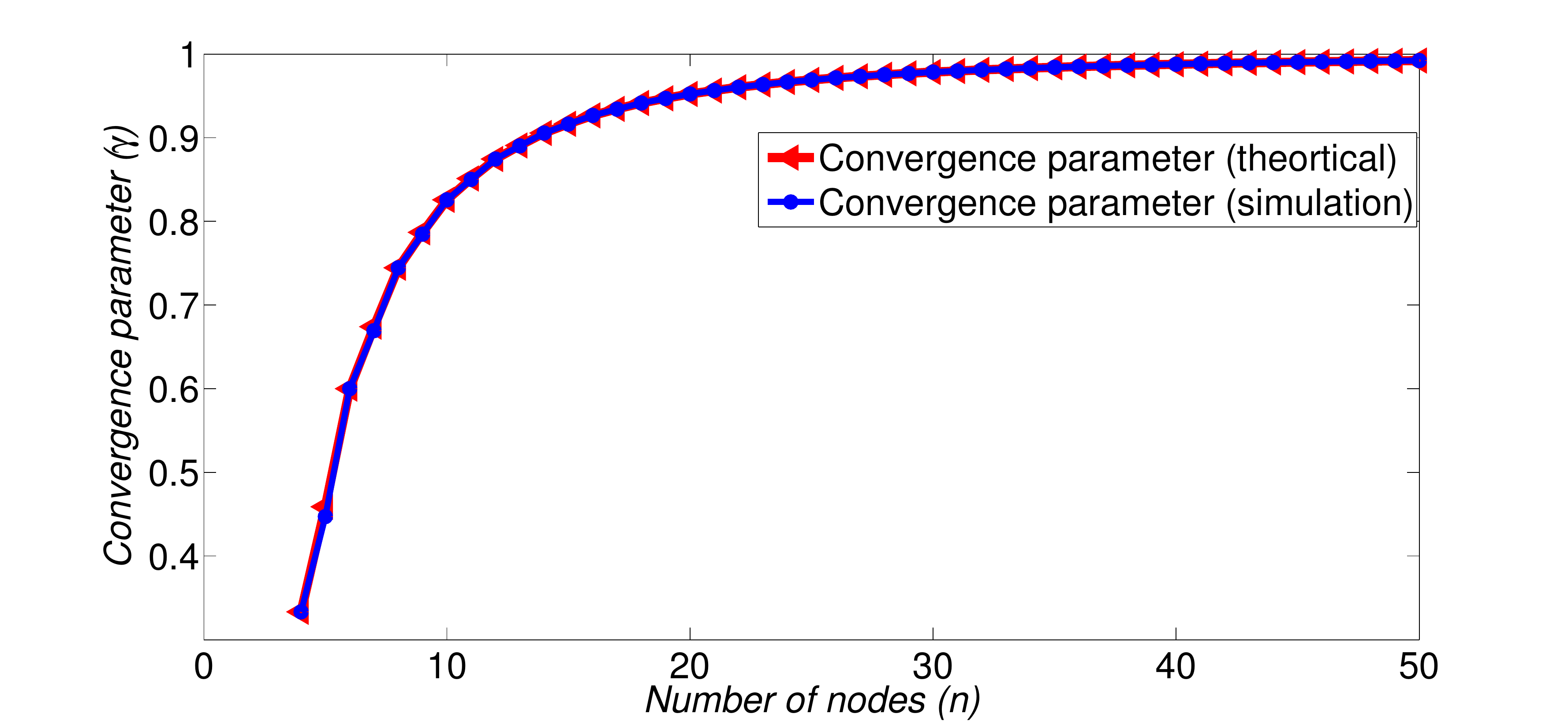}
\caption{Comparison of theoretical and simulation results for convergence parameter $\gamma$ of $1$-nearest neighbor cycle.}
\label{fig:5}
\end{figure}
\begin{figure}[!t]
\centering
\includegraphics[width=8 cm, height=4 cm]{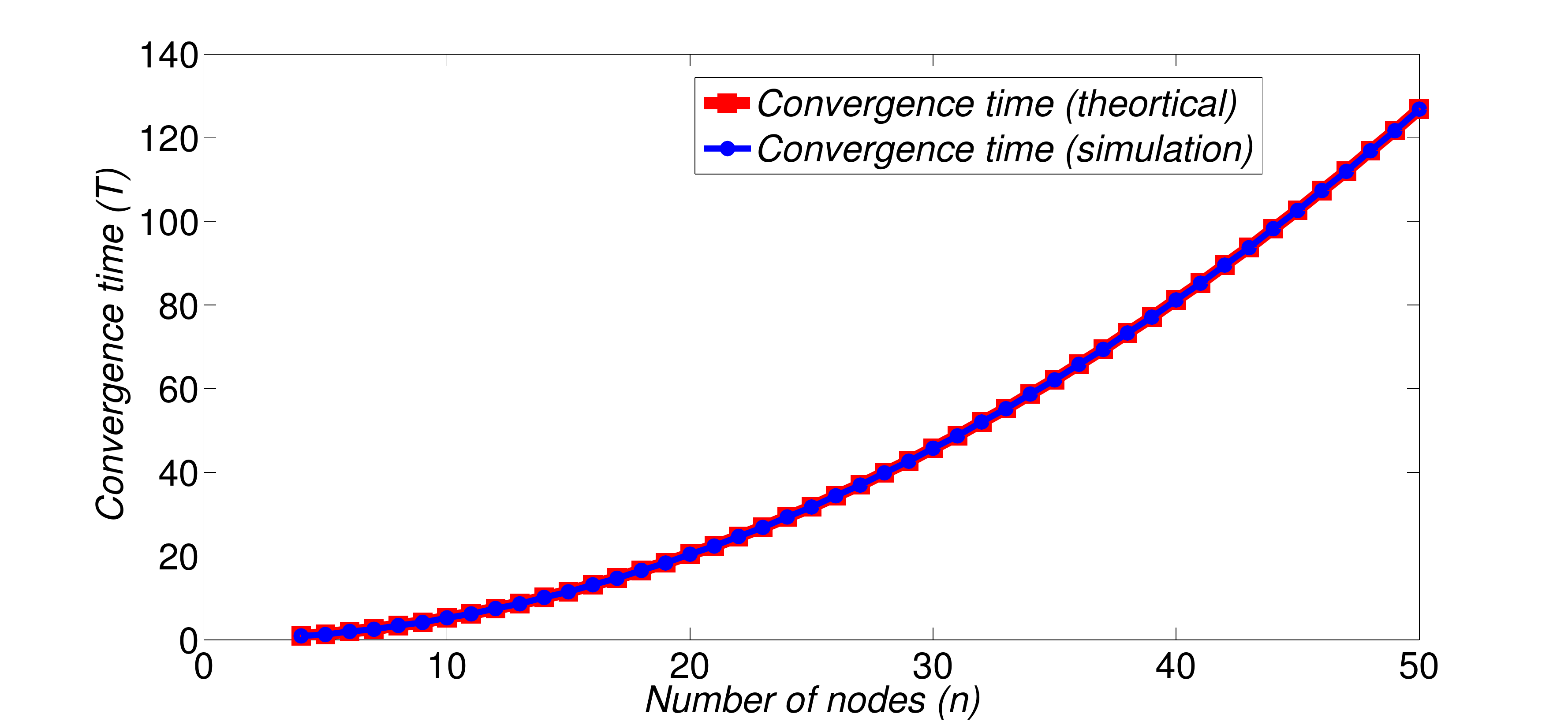}
\caption{Comparison of theoretical and simulation results for convergence time $T$ of $1$-nearest neighbor cycle.}
\label{fig:6}
\end{figure}
\begin{figure}[!t]
\centering
\includegraphics[width=8 cm, height=4 cm]{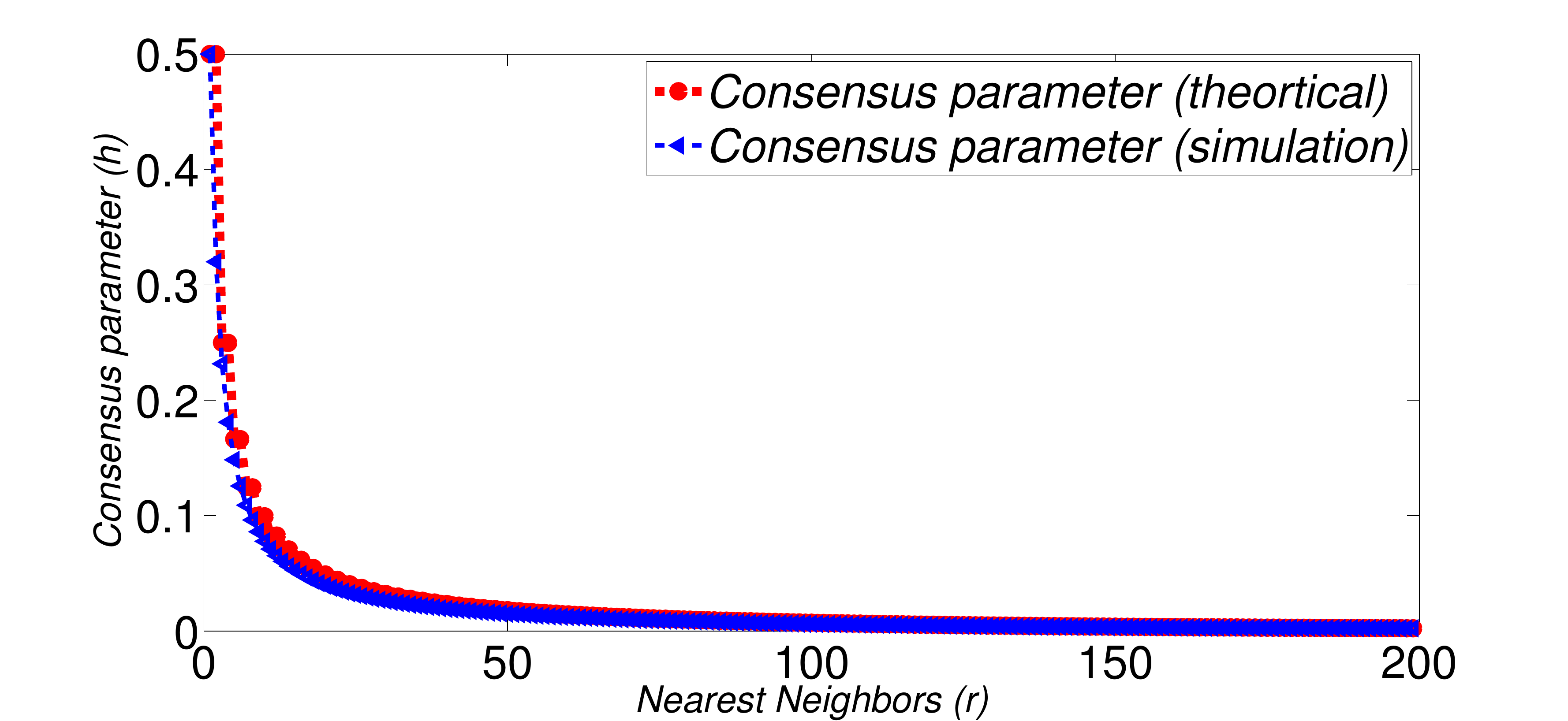}
\caption{Comparison of theoretical and simulation results for consensus parameter $h$ of $r$-nearest neighbor cycle for n=400.}
\label{fig:8}
\end{figure}
\begin{figure}[!t]
\centering
\includegraphics[width=8 cm, height=4 cm]{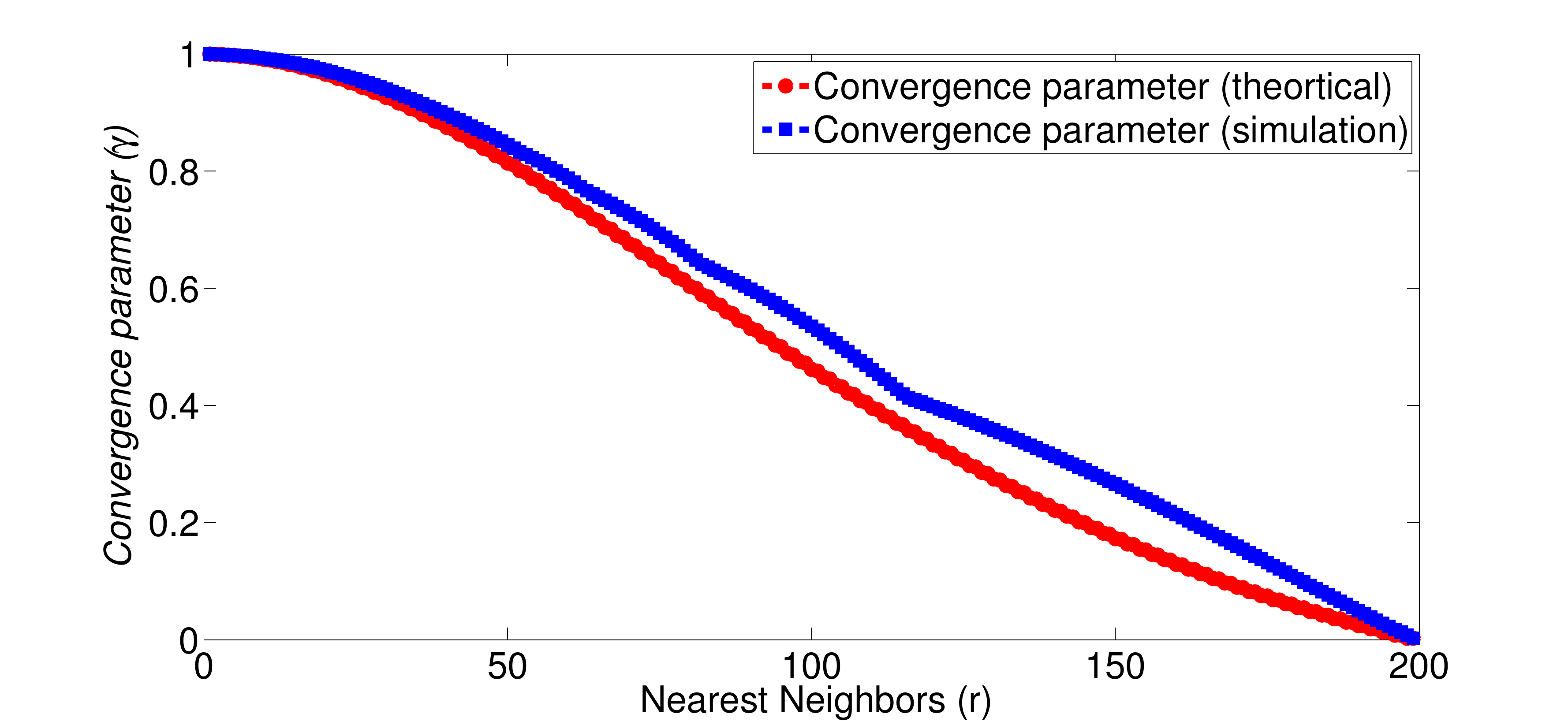}
\caption{Comparison of theoretical and simulation results for convergence parameter $\gamma$ of $r$-nearest neighbor cycle for n=400.}
\label{fig:9}
\end{figure}
\begin{figure}[!t]
\centering
\includegraphics[width=8 cm, height=4 cm]{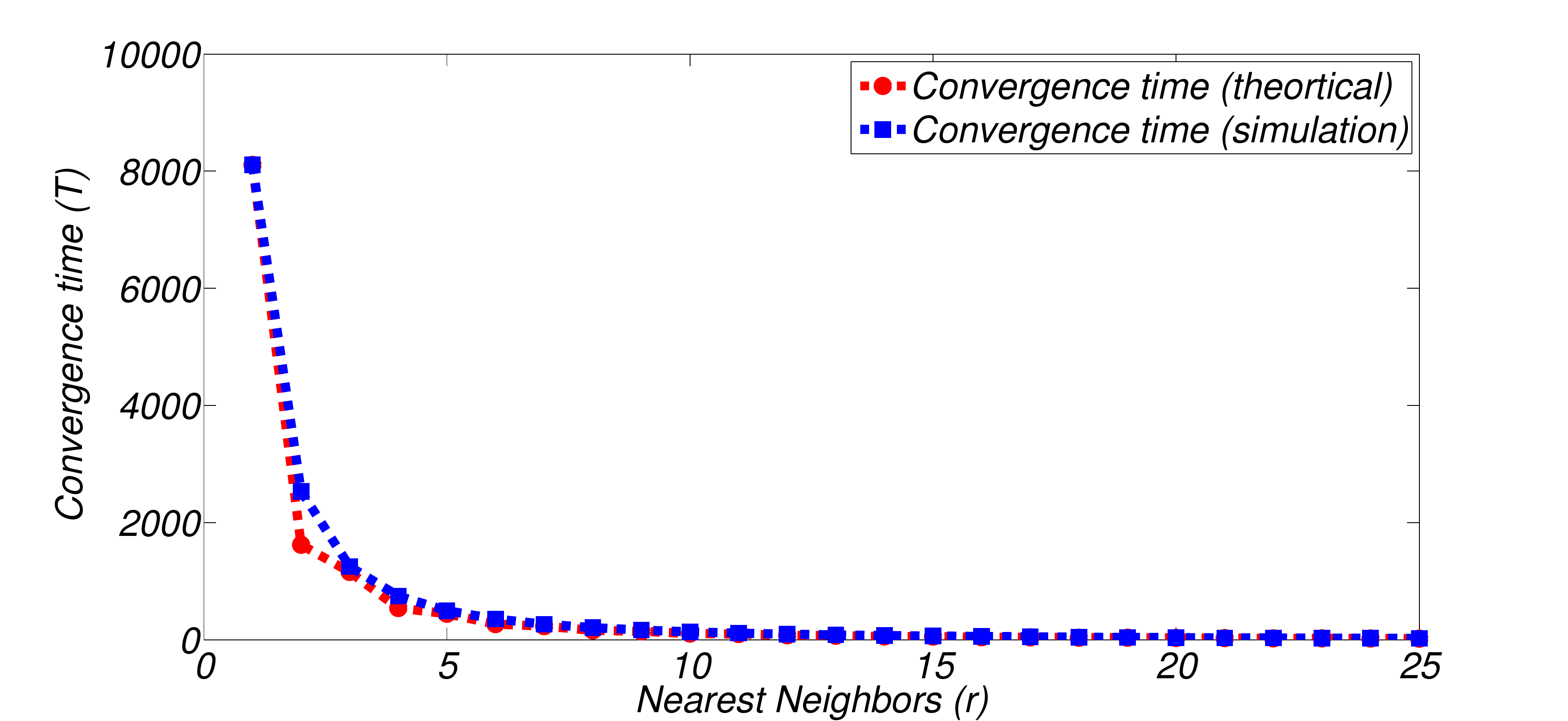}
\caption{Comparison of theoretical and simulation results for convergence time $T$ of $r$-nearest neighbor cycle for n=400.}
\label{fig:7}
\end{figure}
\begin{figure}[!t]
\centering
\includegraphics[width=8 cm, height=4 cm]{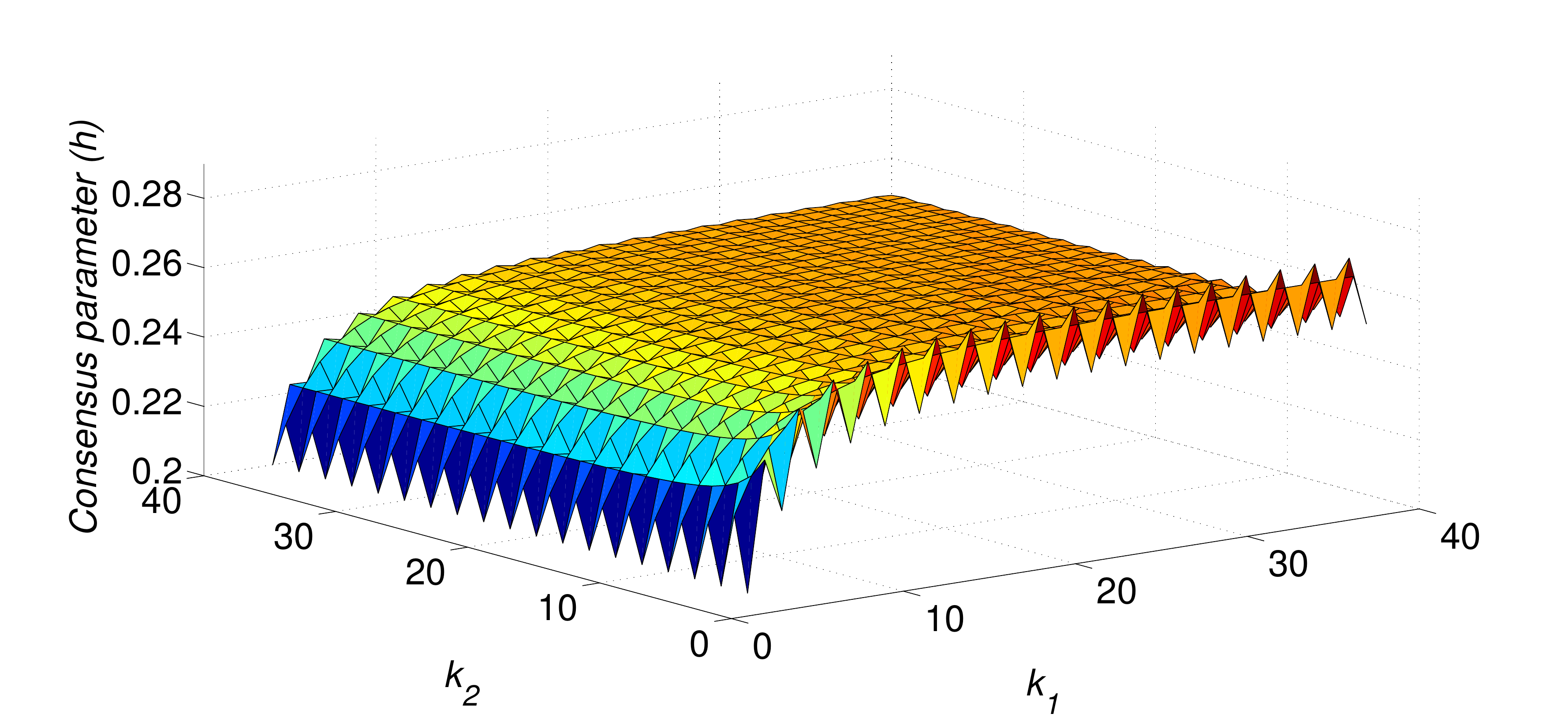}
\caption{Consensus parameter $h$ versus $k_1$, $k_2$ of $r$-nearest neighbor torus for $r=1$.}
\label{fig:11}
\end{figure}
\begin{figure}[!t]
\centering
\includegraphics[width=8 cm, height=4 cm]{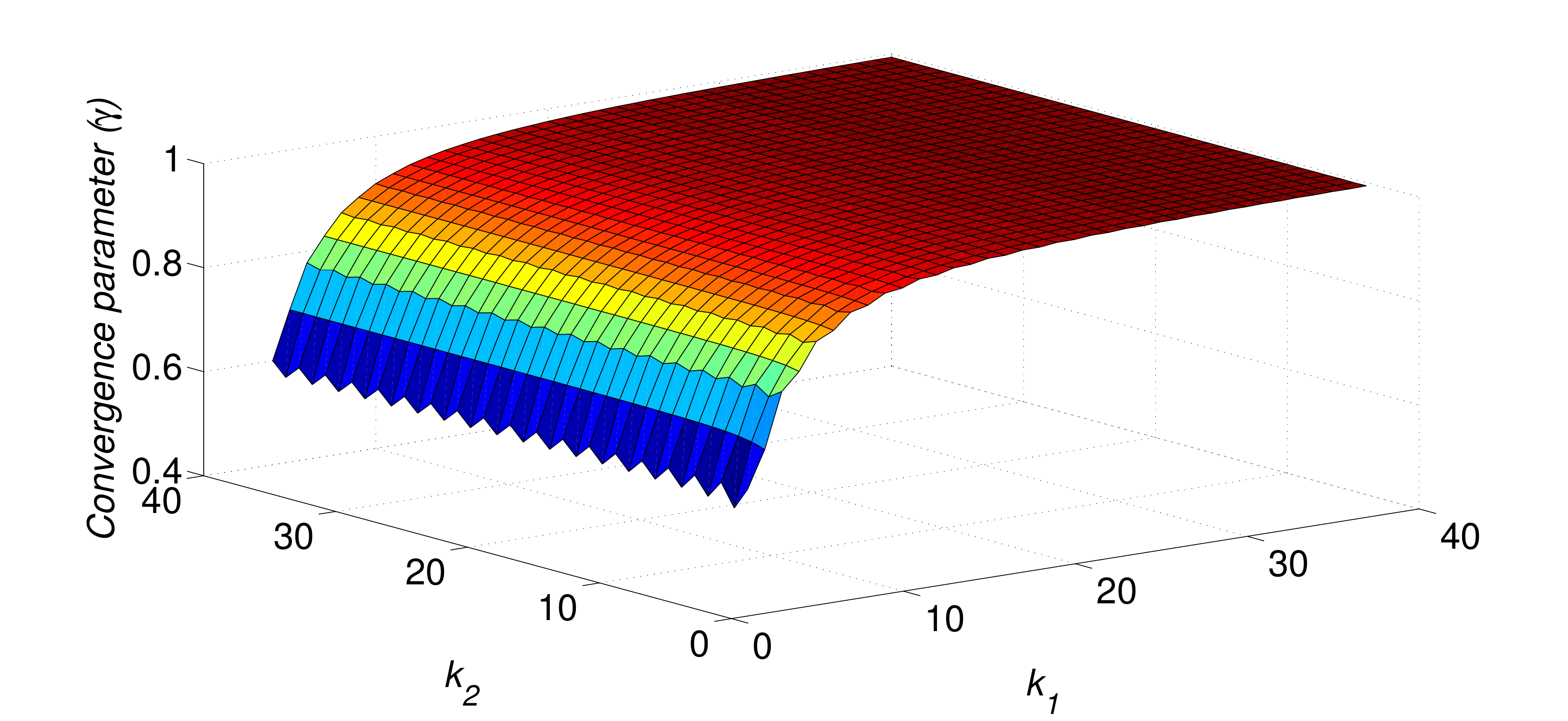}
\caption{Convergence parameter $\gamma$ versus $k_1$, $k_2$ of $r$-nearest neighbor torus for $r=1$.}
\label{fig:12}
\end{figure}
\begin{figure}[!t]
\centering
\includegraphics[width=8 cm, height=4 cm]{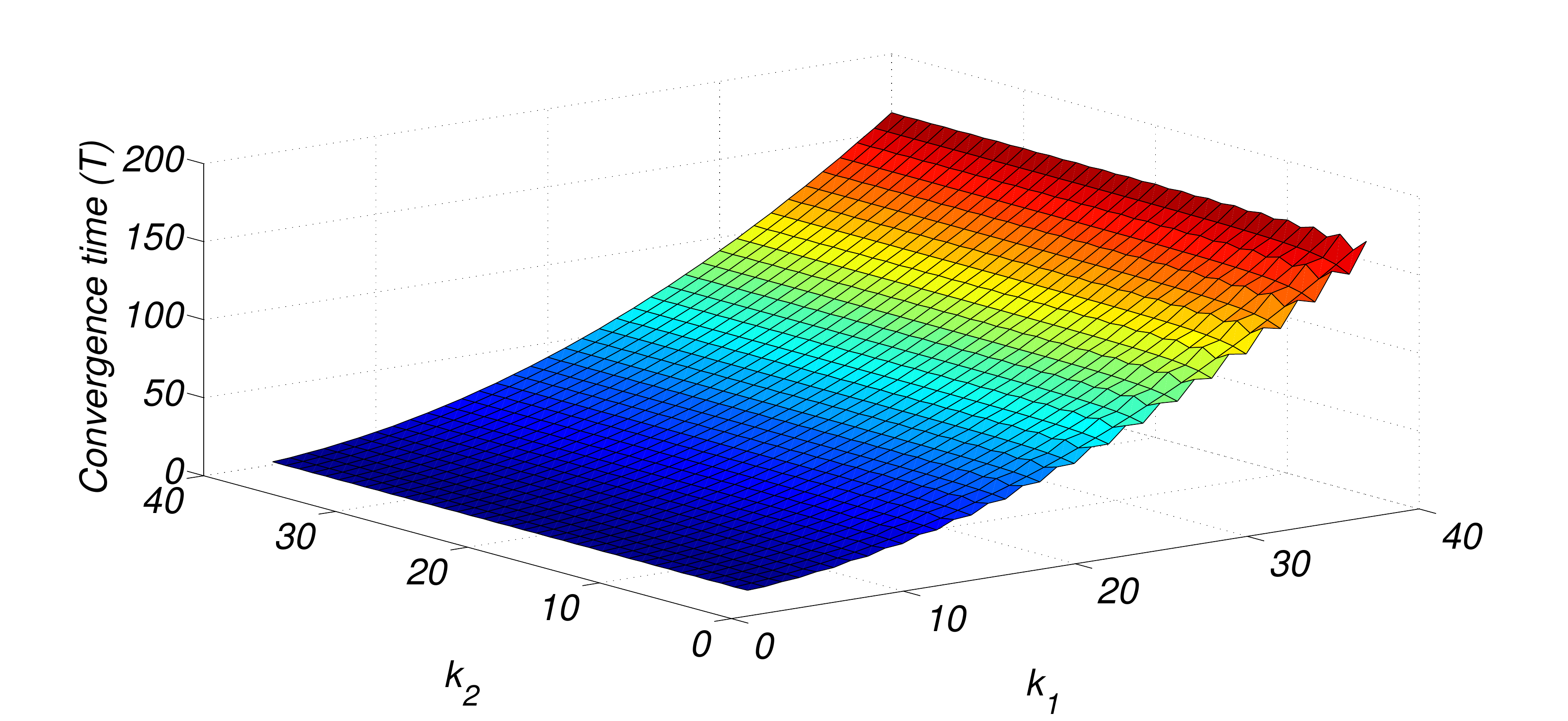}
\caption{Convergence time $T$ versus $k_1$, $k_2$ of $r$-nearest neighbor torus for $r=1$.}
\label{fig:10}
\end{figure}
\begin{figure}[!t]
\centering
\includegraphics[width=8 cm, height=4 cm]{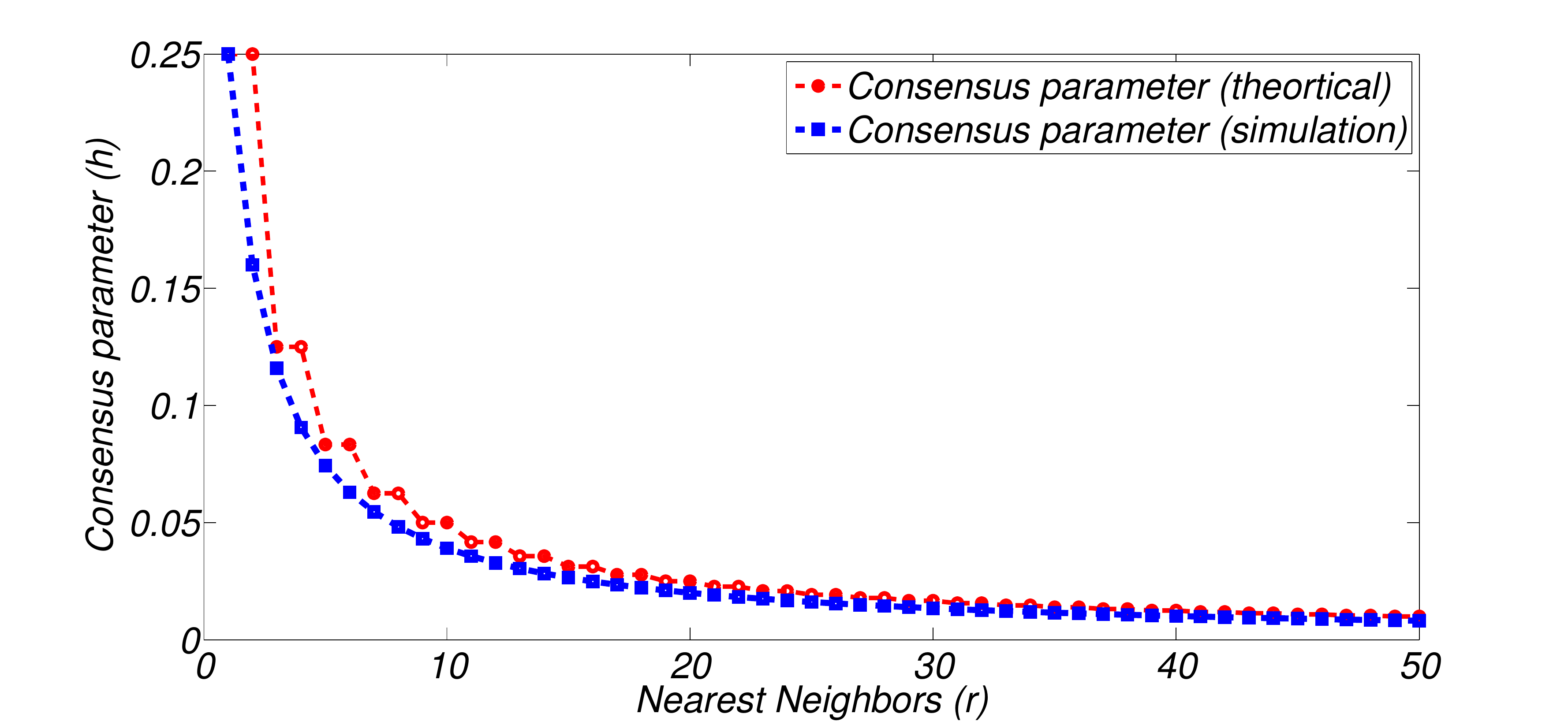}
\caption{Comparison of theoretical and simulation results for consensus parameter $h$ of $r$-nearest neighbor torus for $k_1=k_2=1000$.}
\label{fig:14}
\end{figure}
\begin{figure}[!t]
\centering
\includegraphics[width=8 cm, height=4 cm]{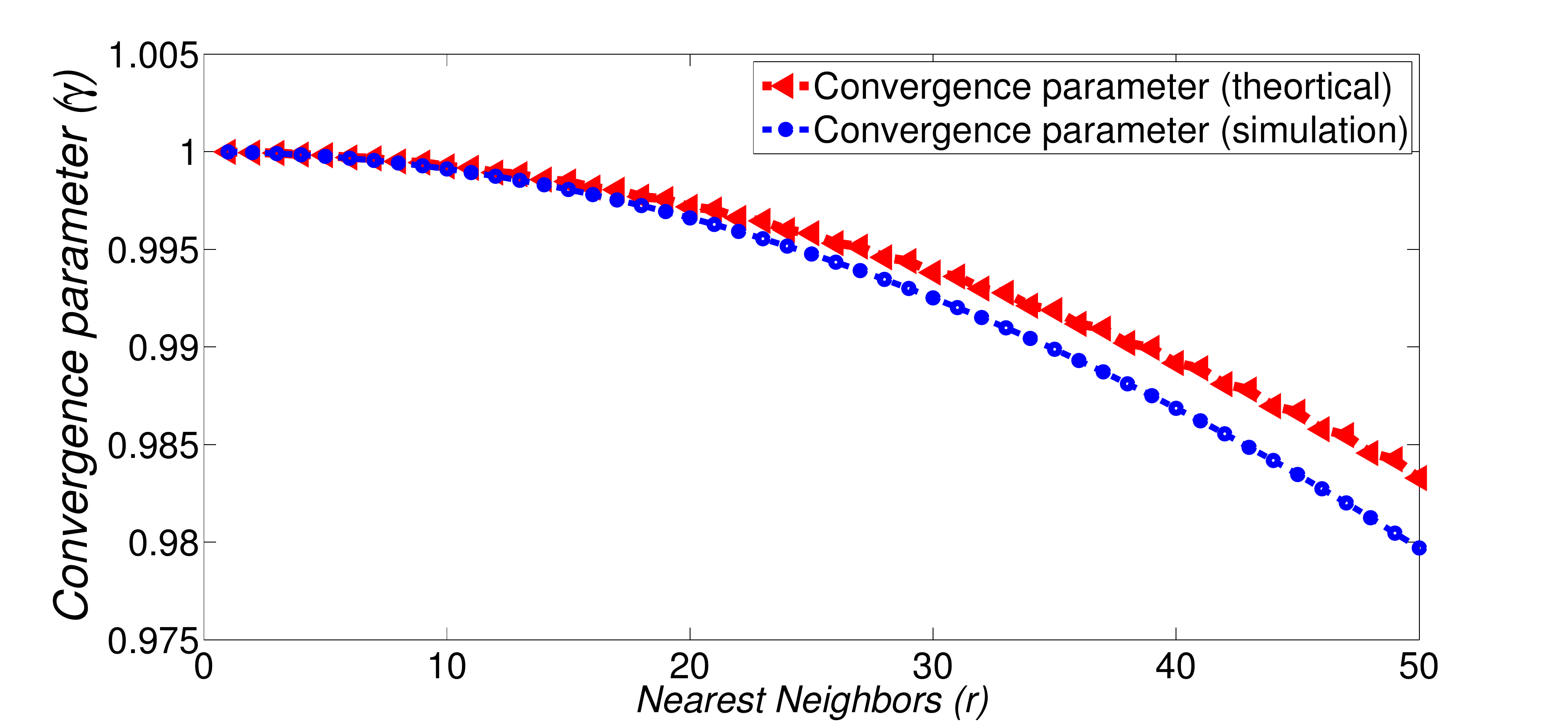}
\caption{Comparison of theoretical and simulation results for convergence parameter $\gamma$ of $r$-nearest neighbor torus for $k_1=k_2=1000$.}
\label{fig:15}
\end{figure}
\begin{figure}[!t]
\centering
\includegraphics[width=8 cm, height=4 cm]{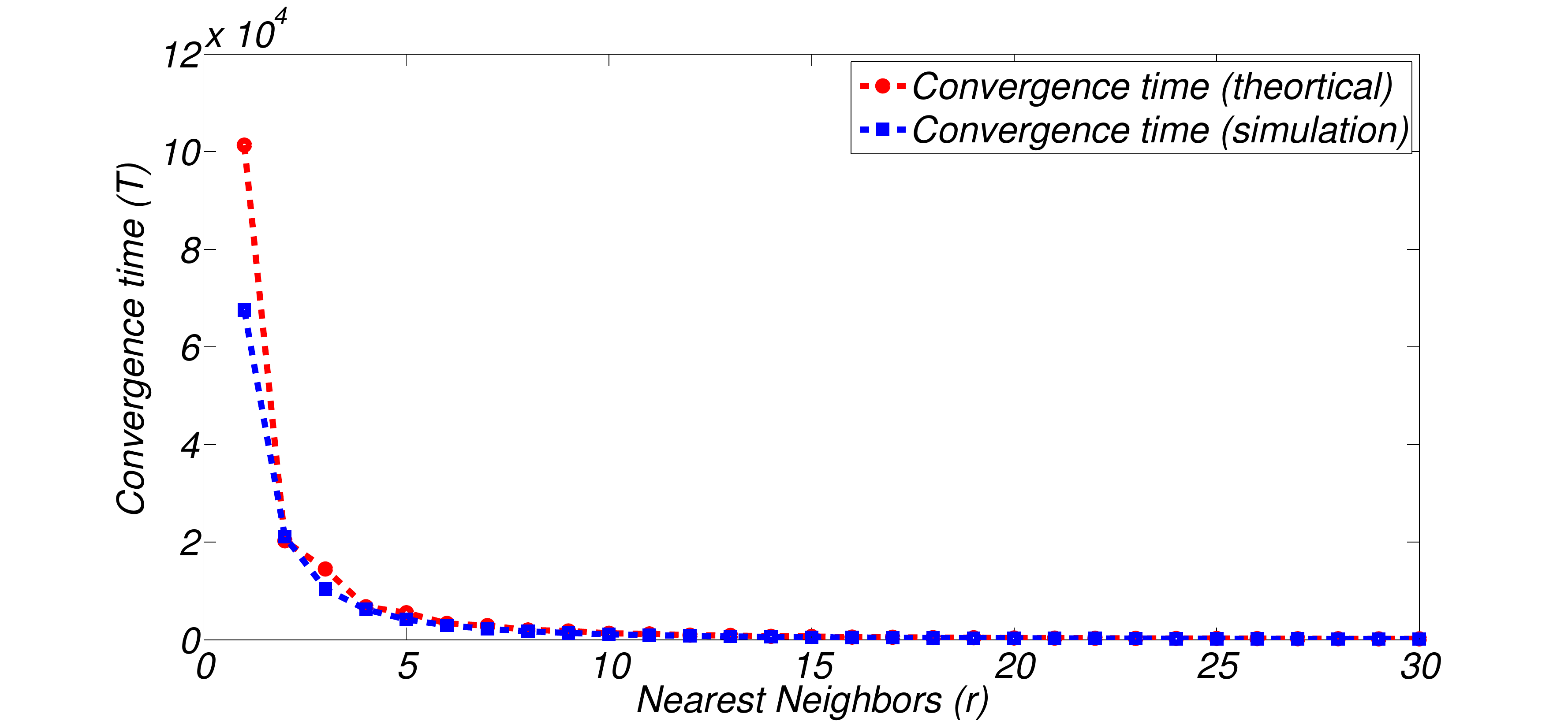}
\caption{Comparison of theoretical and simulation results for convergence time $T$ of $r$-nearest neighbor torus for $k_1=k_2=1000$.}
\label{fig:13}
\end{figure}
\begin{figure}[!t]
\centering
\includegraphics[width=8 cm, height=4 cm]{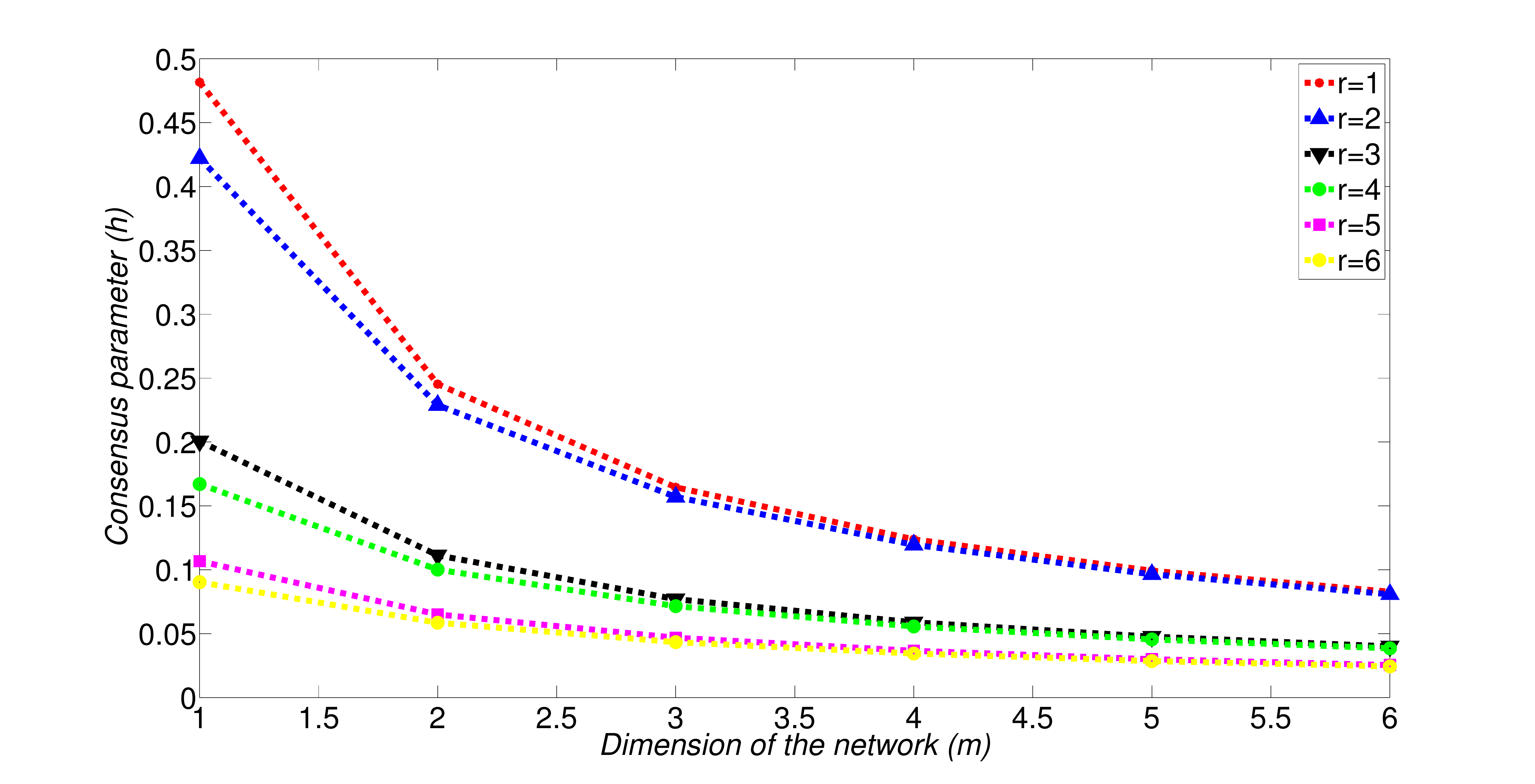}
\caption{Consensus parameter $h$ versus network dimension $m$ of $m$-dimensional torus with
varying number of nearest neighbors $r$ for $k_{1}$ = 16, $k_{2}$ = 18, $k_{3}$ = 20, $k_{4}$ = 22, $k_{5}$ = 24, and $k_{6}$ = 26.}
\label{fig:18}
\end{figure}
\begin{figure}[!t]
\centering
\includegraphics[width=8 cm, height=4 cm]{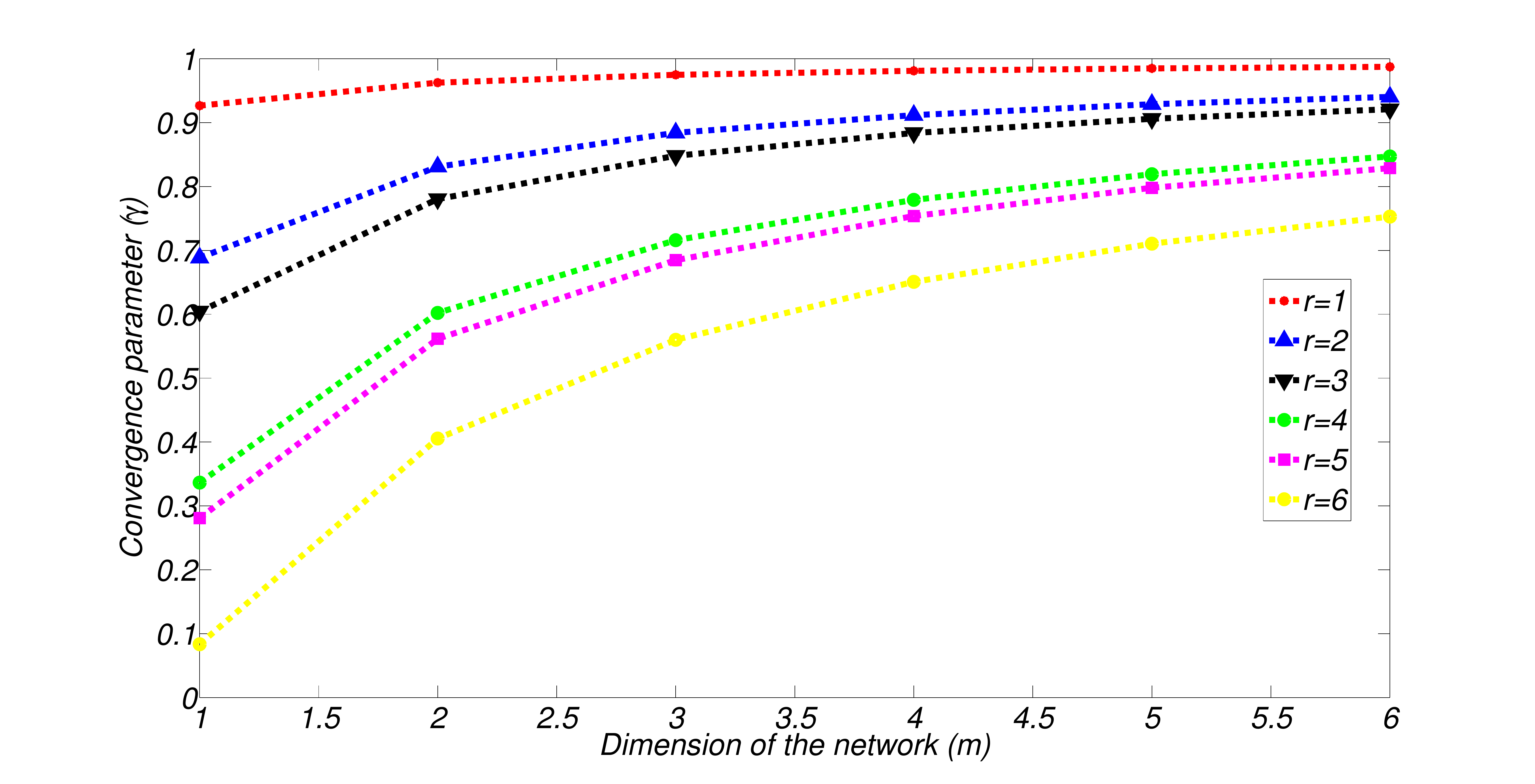}
\caption{Convergence parameter $\gamma$ versus network dimension $m$ of $m$-dimensional torus with
varying number of nearest neighbors $r$ for $k_{1}$ = 16, $k_{2}$ = 18, $k_{3}$ = 20, $k_{4}$ = 22, $k_{5}$ = 24, and $k_{6}$ = 26.}
\label{fig:17}
\end{figure}
\begin{figure}[!t]
\centering
\includegraphics[width=8 cm, height=4 cm]{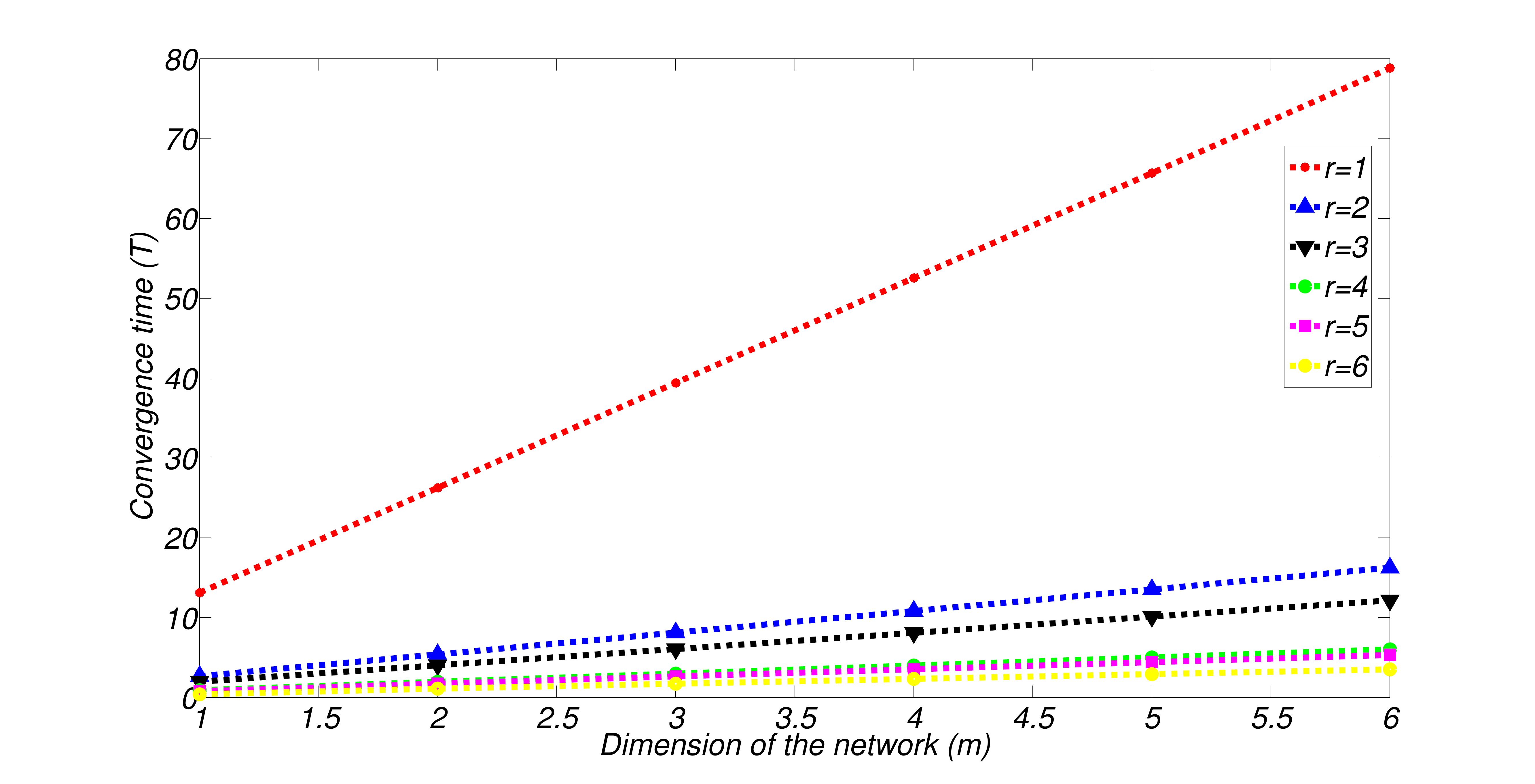}
\caption{Convergence time $T$ versus network dimension $m$ of $m$-dimensional torus with
varying number of nearest neighbors $r$ for $k_{1}$ = 16, $k_{2}$ = 18, $k_{3}$ = 20, $k_{4}$ = 22, $k_{5}$ = 24, and $k_{6}$ = 26.}
\label{fig:16}
\end{figure}
\section{Conclusions}
In this paper, we have derived the analytical expressions for optimal consensus parameter, optimal convergence parameter to estimate the convergence time of $m$-dimensional WSNs. We have investigated that nodes in multidimensional WSNs require more nearest neighbors or large transmission radius without effecting the power consumption. We have also proposed a optimization framework to design and control the performance of consensus algorithm on WSNs. Furthermore, the analytical expressions derived in this paper are extremely useful to exactly estimate the convergence time for large WSNs with less computational complexity.
\appendices
\section{$r$-nearest neighbor cycle}
Proofs of $Theorem$ $6$, $Theorem$ $7$, $Theorem$ $8$, and $Theorem$ $9$ are given below.\\
From $Theorem$ $2$, eigenvalue expressions of weight matrix $W$ can be written as
\begin{equation}
\lambda _0 (W) = \left( {1 - 2rh} \right) + 2h,
\label{51}
\end{equation}
\begin{equation}
\lambda _1 (W) = \left( {1 - 2rh} \right) + 2h\sum\limits_{j = 1}^r {\cos \left( {\frac{{2\pi j}}{n}} \right)},
\label{52}
\end{equation}
\begin{equation}
\lambda _{\frac{n}{2}} (W) = \left( {1 - 2rh} \right) + 2h\sum\limits_{j = 1}^r {\cos \left( {\pi j} \right)},
\label{53}
\end{equation}
\begin{equation}
\lambda _{\frac{{(n - 1)}}{2}} (W) = (1 - 2hr) + 2h\sum\limits_{i = 1}^r {\cos \left( {\frac{{\pi i(n - 1)}}{n}} \right)},
\label{54}
\end{equation}
Note that $\lambda _1(W)$ is a second largest eigenvalue and $\lambda _{\frac{n}{2}}(W)$ is a smallest eigenvalue of $W$, where $n$ is even integer. And $\gamma$ is minimum, when
\begin{equation}
\left| {\lambda _1 (W)} \right| = \left| {\lambda _{\frac{n}{2}} (W)} \right|.
\label{55}
\end{equation}
Substitution of (\ref{52}) and (\ref{53}) in (\ref{55}), results in
\begin{equation}
h  = \frac{1}{{2r+0.5 - \sum\limits_{j = 1}^r {\cos \left( {\frac{{2\pi j}}{n}} \right)}  - \sum\limits_{j = 1}^r {\cos \left( {\pi j} \right)} }}.
\label{56}
\end{equation}
$Note$ 1: Dirichlet kernel is expressed as
\begin{equation}
1 + 2\sum\limits_{j = 1}^r {\cos \left( {jx} \right)}  = \frac{{\sin \left( {r + \frac{1}{2}} \right)x}}{{\sin \frac{x}{2}}}.
\label{57}
\end{equation}
Using (\ref{57}), $h$ can be rewritten as (\ref{36}). Finally, substitution of (\ref{36}) in (\ref{52}) proves $Theorem$ $8$. \\
Since $\lambda _{\frac{n-1}{2}}(W)$ is a smallest eigenvalue, where $n$ is odd integer, and $\gamma$ is minimum, when 
\begin{equation}
\left| {\lambda _1 (W)} \right| = \left| {\lambda _{\frac{{(n - 1)}}{2}} (W)} \right|.
\label{58}
\end{equation}
Substitution of (\ref{52}) and (\ref{54}) in (\ref{58}) results in
\begin{equation}
h  = \frac{1}{{2r - \sum\limits_{j = 1}^r {\cos \left( {\frac{{2\pi j}}{n}} \right)}  - \sum\limits_{j = 1}^r {\cos \left( {\frac{{\pi j\left( {n - 2} \right)}}{n}} \right)} }},
\label{59}
\end{equation}
So from (\ref{57}), $h$ can be further simplified as (\ref{37}). Finally, substitution of (\ref{37}) in (\ref{52}), proves $Theorem$ $9$.
\section{$r$-nearest neighbor two dimensional torus}
Proofs of $Theorem$ $10$, $Theorem$ $11$, $Theorem$ $12$, and $Theorem$ $13$ are given below. \\
From $Theorem$ $4$, eigenvalue expressions of $W$ can be written as
\begin{equation}
\lambda _{0,1} (W) = (1 - 2hr) + 2h\sum\limits_{j = 1}^r {\cos \left( {\frac{{2\pi j}}{{k_2 }}} \right)}, 
\label{60}
\end{equation}
\begin{equation}
\lambda _{\frac{k_1}{2},\frac{k_2}{2}} (W) = (1 - 4hr) + 4h\sum\limits_{j = 1}^r {\cos \left( {\pi j} \right)},
\label{61}
\end{equation}
\begin{equation}
\resizebox{.9 \hsize} {!} {$\lambda _{\frac{{\left( {k_1 - 1} \right)}}{2},\frac{{\left( {k_2 - 1} \right)}}{2}} (W) = (1 - 4hr) + 2h\sum\limits_{j = 1}^r {\cos \left( {\frac{{\pi j\left( {k_1  - 1} \right)}}{{k_1 }}} \right)}  + 2h\sum\limits_{j = 1}^r {\cos \left( {\frac{{\pi j\left( {k_2  - 1} \right)}}{{k_2 }}} \right)}$}.
\label{62}
\end{equation}
We have noticed that $\lambda _{0,1}(W)$ is a second largest eigen value of $W$ and $ \lambda _{\frac{{k_1 }}{2},\frac{{k_2 }}{2}} (W)$ is a smallest eigenvalues of $W$, when $k_1, k_2$ are even integers. And $\gamma$ is minimum, when
\begin{equation}
\left| {\lambda _{0,1} (W)} \right| = \left| {\lambda _{\frac{{k_1 }}{2},\frac{{k_2 }}{2}} (W)} \right|.
\label{63}
\end{equation}
Substitution of (\ref{60}) and (\ref{61}) in (\ref{63}) results in
\begin{equation}
h = \frac{1}{{3r - \sum\limits_{i = 1}^r {\cos \left( {\frac{{2\pi i}}{{k_2 }}} \right) - 2\sum\limits_{i = 1}^r {\cos \left( {\pi i} \right)} } }}, 
\label{64}
\end{equation}
Using (\ref{57}), $h$ can be simplified as (\ref{40}). Finally, substitution of (\ref{40}) in (\ref{60}) proves $Theorem$ $12$.\\
It can be easily observed that $\lambda _{\frac{{(k_1-1) }}{2},\frac{{(k_2-1)}}{2}}(W)$ is a smallest eigenvalue when $k_1, k_2$ are odd integers. So $\gamma$ is minimum when,
\begin{equation}
\left| {\lambda _{0,1} (W)} \right| = \left| {\lambda _{\frac{{(k_1  - 1)}}{2},\frac{{(k_2  - 1)}}{2}} (W)} \right|. 
\label{65}
\end{equation}
Substitution of (\ref{60}) and (\ref{62}) in (\ref{65}) results in 
\begin{equation}
 \resizebox{.9 \hsize} {!} {$ h = \frac{1}{{3r - \sum\limits_{i = 1}^r {\cos \left( {\frac{{2\pi i}}{{k_2 }}} \right) - \sum\limits_{i = 1}^r {\cos \left( {\frac{{\pi i(k_1  - 1)}}{{k_1 }}} \right)}  - \sum\limits_{i = 1}^r {\cos \left( {\frac{{\pi i(k_2  - 1)}}{{k_2 }}} \right)} } }}$}.
  \label{66}
 \end{equation}
Using (\ref{57}), $h$ can be further simplified as (\ref{41}). Finally, $Theorem$ $13$ can be proved, by substituting the (\ref{41}) in (\ref{60}).
\section{$m$-dimensional $r$-nearest neighbor torus}
Proofs of $Theorem$ $14$, $Theorem$ $15$, $Theorem$ $16$, and $Theorem$ $17$ are given below.\\
Using $Theorem$ $5$, eigenvalue expressions of weight matrix $W$ can be expressed as
\begin{equation}
\resizebox{.9 \hsize} {!} {$\lambda _{1,0,0.......0} (W) = (1 - 2hr) + 2h\sum\limits_{j = 1}^r {\cos \left( {\frac{{2\pi j}}{{k_1 }}} \right)}$}, 
\label{67}
\end{equation}
\begin{equation}
\resizebox{.9 \hsize} {!} {$\lambda _{\frac{{k_1 }}{2},\frac{{k_2 }}{2}....\frac{{k_m }}{2}} (W) = (1 - 2mhr) + 2mh\sum\limits_{j = 1}^r {\cos \left( {\pi j} \right)}$},
\label{68}
\end{equation}
\begin{equation}
\resizebox{.9 \hsize} {!} {$\lambda _{\frac{{\left( {k_1  - 1} \right)}}{2},\frac{{\left( {k_2  - 1} \right)}}{2}....\frac{{\left( {k_m  - 1} \right)}}{2}} (W) = (1 - 2mhr) + 2h\sum\limits_{i = 1}^m {\sum\limits_{j = 1}^r {\cos \left( {\frac{{\pi j\left( {k_i  - 1} \right)}}{{k_i }}} \right)} }$}.
\label{69}
\end{equation}
Note that $\lambda _{1,0,0..........0} (W) $ is a second largest eigenvalue and  $\lambda _{\frac{{k_1 }}{2},\frac{{k_2 }}{2},..........\frac{{k_m }}{2}} (W) $ is a smallest eigenvalue of weight matrix $W$, when $k_1, k_2, k_3...k_m$ are even integers. And $\gamma$ is minimum when
\begin{equation}
\left| {\lambda _{1,0,0......0} (W)} \right| = \left| {\lambda _{\frac{{k_1 }}{2},\frac{{k_2 }}{2},........\frac{{k_m }}{2}} (W)} \right|.
\label{70}
\end{equation}
Substitution of (\ref{67}) and (\ref{68}) in (\ref{70}) results in
\begin{equation}
h = \frac{1}{{r(m + 1) - \sum\limits_{i = 1}^r {\cos \left( {\frac{{2\pi i}}{{k_1 }}} \right) - m\sum\limits_{i = 1}^r {\cos \left( {\pi i} \right)} } }}.
\label{71}
\end{equation}
Using (\ref{57}), $h$ can be further simplified as (\ref{44}). $Theorem$ $16$ can be proved by substituting the (\ref{44}) in (\ref{67}).\\
We have noticed that, $\lambda _{1,0,0..........0} (W) $ is a second largest eigenvalue and $\lambda _{\frac{{k_1-1 }}{2},\frac{{k_2-1 }}{2},..........\frac{{k_m-1}}{2}} (W) $ is a smallest eigenvalues of weight matrix $W$, when $k_1, k_2, k_3...k_m$ are odd integers. And $\gamma$ is minimum when 
 \begin{equation}
\left| {\lambda _{1,0,0.......0} (W)} \right| = \left| {\lambda _{\frac{{(k_1  - 1)}}{2},\frac{{(k_2  - 1)}}{2},.......,\frac{{(k_m  - 1)}}{2}} (W)} \right|.
\label{72}
\end{equation}
Finally, substitution of (\ref{67}) and (\ref{69}) in (\ref{72}) results in
\begin{equation}
 h = \frac{1}{{r(m + 1) - \sum\limits_{i = 1}^r {\cos \left( {\frac{{2\pi i}}{{k_1 }}} \right) - \sum\limits_{j = 1}^m {\sum\limits_{i = 1}^r {\cos \left( {\frac{{\pi i(k_j  - 1)}}{{k_j }}} \right)} } } }}.
 \label{73}
 \end{equation}
Using (\ref{57}), $h$ can be further simplified as (\ref{45}). Finally, substitution of (\ref{45}) in (\ref{67}) proves the $Theorem$ $17$.


\ifCLASSOPTIONcaptionsoff
  \newpage
\fi



%

%
\bibliographystyle{IEEEtran}
\bibliography{papers}
\begin{IEEEbiographynophoto}{Sateeshkrishna Dhuli}
received his M. Tech. degree in Electronics and communication 
Engineering from National Institute of Technology, Warangal, India in 2011 and B. Tech. degree in Electronics and communication 
Engineering from  Maharaj Vijayaram Gajapathi Raj College of Engineering, Vizianagaram, India in 2008. He is currently  working  as  Doctoral student at Electrical  Engineering, Indian Institute of Technology,  Kanpur,  India. His research interest includes
Wireless sensor networks, Complex networks, and Mobile adhoc networks.
\end{IEEEbiographynophoto}
\begin{IEEEbiographynophoto}{Kumar gaurav} received his B. Tech. degree in Electronics and Telecommunication 
Engineering from College of Engineering Roorkee, India in 2011. He is currently  working as Doctoral student at Electrical  Engineering, Indian Institute  of  Technology, Kanpur, India. His research interest includes Wireless networks, Social networks, and Biological networks.
\end{IEEEbiographynophoto}
\begin{IEEEbiographynophoto}{Y.N.Singh} received  his  Ph. D. and  M. Tech. degrees  in  Electrical
Engineering from Indian Institute of Technology, Delhi, India in 1997
and 1992, respectively, and B. Tech. degree in Electrical Engineering
from  Regional  Engineering  College, Hamirpur, India in 1991. He is currently  working  as  Professor at Electrical  Engineering, Indian  Institute  of  Technology, Kanpur, India. His research interest includes
Optical Communication Networks, Mobile Adhoc Networks, and Computer Networks. Dr. Singh has published papers in International
Journals and Conferences including IEEE etc.
\end{IEEEbiographynophoto}

%
%




\end{document}